# Formal Metatheory of Second-Order Abstract Syntax

MARCELO FIORE* and DMITRIJ SZAMOZVANCEV†, University of Cambridge, UK

Despite extensive research both on the theoretical and practical fronts, formalising, reasoning about, and implementing languages with variable binding is still a daunting endeavour – repetitive boilerplate and the overly complicated metatheory of capture-avoiding substitution often get in the way of progressing on to the actually interesting properties of a language. Existing developments offer some relief, however at the expense of inconvenient and error-prone term encodings and lack of formal foundations.

We present a mathematically-inspired language-formalisation framework implemented in Agda. The system translates the description of a syntax signature with variable-binding operators into an intrinsically-encoded, inductive data type equipped with syntactic operations such as weakening and substitution, along with their correctness properties. The generated metatheory further incorporates metavariables and their associated operation of metasubstitution, which enables second-order equational/rewriting reasoning. The underlying mathematical foundation of the framework – initial algebra semantics – derives compositional interpretations of languages into their models satisfying the semantic substitution lemma by construction.

## 1 INTRODUCTION

Programming and logic research papers that introduce and study new languages and calculi with variable binding typically gloss over the associated notion of capture-avoiding substitution – it is often taken as standard along with its correctness properties. Nevertheless, the representation of variables and substitution becomes a major roadblock when one attempts to formalise such languages in proof assistants, as substituting into terms with variable binders involves dealing with variable shadowing and capture. Some ways of encoding variables (e.g. using strings or numeric de Bruijn indices) lead to fragile and error-prone implementations of syntactic operations: for instance, binding links are easily broken if one increments the wrong index or forgets to modify a string. Other approaches – such as type- and scope-safe de Bruijn indices – place static guarantees on the correctness of the implementation but require significant boilerplate and seemingly superfluous generality that puts the interesting metatheoretic proofs on hold. Either way, the end result is a formalisation littered with ad hoc definitions and lemmas about weakening, substitution, etc., all of which need to be tweaked if the syntax is changed even slightly.

We tackle this unsatisfactory state of affairs by stepping back and considering the mathematical foundations of *second-order abstract syntax*; that is, abstract syntax with variable binding and parametrised metavariables. Using an approach derived from the presheaf model developed by Fiore, Plotkin, and Turi [1999] and Fiore [2008], we build a powerful and entirely generic framework for language formalisation that extracts the maximum amount of syntactic metatheory from a second-order syntax description with minimal boilerplate and user effort. From a concise textual specification of a typed syntax with binding operators, such as the following one for the simply-typed lambda calculus (STLC)

---

*Research partially supported by EPSRC grant EP/V002309/1.
†Research supported by EPSRC grant EP/R513180/1.

---

Authors' address: Marcelo Fiore, marcelo.fiore@cst.cam.ac.uk; Dmitrij Szamozvancev, ds709@cst.cam.ac.uk, Department of Computer Science and Technology, University of Cambridge, UK.

| type | | term | | |
|---|---|---|---|---|
| N | : 0-ary | app : $\alpha \succ \beta$ | $\alpha$ | $\rightarrow \beta$ |
| $\_\succ\_$ | : 2-ary | lam : $\alpha.\beta$ | | $\rightarrow \alpha \succ \beta$ |

with base type N and function type $\succ$, our system generates Agda code for: (*i*) a grammar of types and an intrinsically-typed data type of terms; (*ii*) operations of weakening and substitution together with their correctness properties; (*iii*) a record that, when instantiated with a mathematical model, induces a semantic interpretation of the syntax in the model that preserves substitution; (*iv*) a term constructor for parametrised metavariables and their associated operation of metasubstitution; and (*v*) an equational/rewriting theory that can be instantiated with the axioms of the syntax to obtain a library for second-order equational/rewriting reasoning.

## 1.1 Background

The framework presented in this paper builds on a long series of practical and theoretical developments in the study of abstract syntax, incorporating elements from both lines of research.

*Intrinsic typing.* Among the numerous language-formalisation strategies available for a variety of proof assistants, the type- and scope-safe, intrinsically-typed encoding of typed syntax with variable binding popularised by Altenkirch and Reus [1999], Benton et al. [2012], and, more recently, by Allais, Atkey, Chapman, McBride, and McKinna [2021] (henceforth cited as AACMM [2021]) stands out as an excellent fit for dependently-typed metalanguages like Agda [Norell 2009] and Coq [The Coq Development Team 2004]. It represents terms of a syntax as a type- and context-indexed family $\mathcal{X}$ of sets: for a type $\alpha$ and typing context (list of types) $\Gamma$, the set $\mathcal{X} \alpha \Gamma$ consists of the terms $t$ that satisfy the typing judgment $\Gamma \vdash t : \alpha$. In an intrinsically-typed formalisation this indexed family $\mathcal{X}$ is generated inductively: for example, application in the syntax of the STLC corresponds to a constructor app that combines a function term $f : \mathcal{X} (\alpha \succ \beta) \Gamma$ and an argument term $a : \mathcal{X} \alpha \Gamma$ into $\text{app}(f, a) : \mathcal{X} \beta \Gamma$. As such, constructors of the syntax are directly encoded as their own typing rules, making ill-typed and ill-scoped terms unrepresentable.

In contrast, an extrinsic encoding resembles the way type systems are presented in the research literature: one has a grammar of raw terms $t$ and an inductively-defined well-typedness relation $\Gamma \vdash t : \alpha$. While both approaches work and are widely used, dependent types and intrinsic encoding complement each other very well. For instance, in the introductory textbook *Programming Language Foundations in Agda*, Kokke, Siek, and Wadler [2020] present the formalisation of the STLC in both styles, clearly highlighting and advocating the superiority of intrinsic typing.

Another advantage of intrinsic typing is that the need for separate type-preservation proofs of syntactic operations is eliminated: for example, defining the single-variable substitution operation $\text{sub}_1 : \mathcal{X} \alpha \Gamma \rightarrow \mathcal{X} \beta (\alpha \cdot \Gamma) \rightarrow \mathcal{X} \beta \Gamma$ immediately establishes the fact that substitution preserves typing. Just as the typing rules of terms are baked into their syntax, the well-typedness proofs of operations are baked into their definition. The downside is that implementing such operations often requires significant effort, with $\text{sub}_1$ being a prime example: it is not definable from first principles, because the induction hypothesis (viz. the recursive call) is not general enough to handle binding terms, whose fresh bound variables get added to the end of the typing context and thereby "cover" the free variable one wants to substitute for. Instead, someone unfamiliar with the approach has to walk through a long and frustrating path of trial and error. For instance: trying to define weakening and exchange (which are usually employed in the pen-and-paper well-typedness proof) is similarly futile; generalising to substitution for a variable in the middle of the context works for binding terms but not for variables; further generalising to simultaneous substitution still requires weakening; single-variable weakening cannot be implemented directly either, and must be derived from variable renaming. Even worse is that reasoning about single-variable substitution (proving the syntactic or semantic substitution lemmas, for example) forces one to prove similar properties about renaming, weakening, and simultaneous substitution, leading to a tedious and bloated formalisation that is specific to a particular syntax.

*Generic traversals.* If we accept the hoops needed to jump through to implement single-variable substitution, we can look for common patterns to abstract out – after all, renaming and substitution both proceed by traversing a term and replacing variables with *data*, either a variable or a term. McBride [2005] builds on this observation to axiomatise the properties that such *data* need possess in order to define a generic term-traversal function that can be instantiated to either renaming or substitution. Allais et al. [2017] generalise traversals to semantic interpretations in models of the syntax, recovering renaming and substitution as interpretations into the syntactic model. They also give a pattern for proving simulation and fusion properties of traversals, which capture – among other things – the correctness properties of renaming and substitution. In AACMM [2021], the authors further generalise their work on the STLC to an arbitrary universe of second-order syntaxes, producing a robust and flexible language-formalisation framework.

*Presheaf model.* The presheaf model of second-order abstract syntax is a category-theoretic approach to enriching typed languages with variable binding by means of substitution and metasubstitution structures, developed by Fiore and collaborators; see e.g. [Fiore et al. 1999; Fiore 2008; Fiore and Hamana 2013; Fiore and Hur 2010; Fiore and Mahmoud 2010]. It too represents syntax as a type- and context-indexed family of sets, but further requires it to have a functorial renaming structure upon which substitution, metasubstitution, etc. structures are built in a systematic categorical way. While it is an elegant and powerful mathematical theory, its adoption in language-formalisation frameworks has been only incidental: several concepts central to the abstract categorical approach (coends, colimits, etc.) are hard to represent in a dependently-typed setting and a faithful reproduction of the theory would need to heavily rely on a formalised category-theory library such as `agda-categories` [Hu and Carette 2021] or UniMath [Voevodsky et al. 2014].

Recent work by Borthelle et al. [2020] proposed that the presheaf model be adapted to a setting of sorted families by axiomatising the renaming structure and its interaction with initial algebra semantics – ideas which have been explored informally by Allais et al. [2017] and Kaiser et al. [2018] as well. To note, however, is that their skew-monoidal [Szlachányi 2012] approach still depends on categorical theory and results that lead to an impractical formalisation: for example, the syntax of terms is presented by means of the colimit of an $\omega$-chain (and therefore represented by equivalence classes), rather than by an inductive data type.

## 1.2 Contributions

Our work aims to place the constructions and empirical observations of intrinsically-typed term encodings on a formal grounding, thereby bridging the gap between theory and practice. We are closely guided by the presheaf model of second-order abstract syntax, adapting and re-working its categorical theory (definitions and results) to a dependently-typed setting without sacrificing practicality and generality over arbitrary second-order signatures.

- In Section 2 we give the basic mathematical framework of sorted families, renaming coalgebras, and substitution monoids employed in the library. These notions are adapted from the presheaf model [Fiore et al. 1999] which, on its own, is too abstract and high-level to be used as a basis for practical language formalisation. However, with this new (and mathematically justifiable) change of perspective, we can translate all the theory of the presheaf approach to type- and context-indexed families and, in doing so, shine a light upon the constructions and properties commonly encountered in type- and scope-safe treatments of syntax.

- Section 3 outlines the main conceptual tool we rely on: initial algebra semantics. Crucially, not only do we invoke initiality to derive generic traversals such as renaming and substitution, we also use it to prove their correctness, including the associativity of substitution and its interaction with renaming. The notion of a $\Sigma$-monoid is introduced; it axiomatises the structure of a model of the syntax where terms can be interpreted in a compositional and substitution-preserving manner. We also discuss metavariables and metasubstitution [Fiore 2008], together with their application to second-order equational/rewriting reasoning [Fiore and Hur 2010].
- After developing the entirely signature-generic metatheory, in Section 4 we present the translation of second-order syntax descriptions to signature endofunctors and give two approaches of encoding the initial algebra of terms for a signature.
- Finally, in Section 5, we showcase the features of our framework in two extended examples: the syntax of the STLC together with its sound denotational and operational semantics, and the equational theory of partial differentiation.

The paper is intended for an audience familiar with dependently-typed proof assistants and the struggles of language formalisation. Even though they play a vital role in the development, we chose to minimise references to advanced categorical concepts. In the listings we use "retouched Agda" which omits inessential implementation details to make the code cleaner, without impacting clarity (for example, we hide some implicit arguments and record declarations, and use shading instead of cong in equational proofs). The full formalisation with extended proofs and examples can be found on the project website.

## 2  MATHEMATICAL FOUNDATIONS

We start by setting up the abstract mathematical foundations of our work, with the aim of decoupling the renaming and substitution structures from a particular syntax of terms. Section 2.1 lists the standard definitions for contexts, families, and variables, highlighting their important categorical properties. In Section 2.2 we introduce the coalgebraic and monoidal views of renaming and substitution, and in Section 2.3 we consider properties of maps parametrised by a substitution mapping which will be important players in the development of initial algebra semantics.

### 2.1  Contexts and Families

The definitions of contexts and variables are well-known from intrinsically-typed treatments of syntax. Rather than using named variable-type pairs, contexts are lists of types that come from a fixed set $T$, and variables are typed and scoped de Bruijn indices into the context: new points to the first element of the context, while $old(v)$ points to the variable $v$ in an extended context.

```
data Ctx : Set where                         data I : T → Ctx → Set where
   ∅   : Ctx                                     new :          I α (α · Γ)
   _·_ : (α : T) → (Γ : Ctx) → Ctx              old  : I β Γ → I β (α · Γ)
```

We call context-indexed sets of type $\mathsf{Ctx} \to \mathsf{Set}$ *families*, and sort-indexed families (such as $I$ above) a *sorted family*: for an arbitrary sorted family $\mathcal{X} : T \to \mathsf{Ctx} \to \mathsf{Set}$, an element in the set $\mathcal{X} \, \alpha \, \Gamma$ can be seen as an $\mathcal{X}$-term of type $\alpha$ with free variables in context $\Gamma$. Maps between sorted families are sort- and context-indexed families of functions, and together they form the category $\mathbf{Fam}_s$.

```
Family : Set₁           Familyₛ : Set₁              _⇾_ : Familyₛ → Familyₛ → Set
Family = Ctx → Set      Familyₛ = T → Family        𝒳 ⇾ 𝒴 = {α : T}{Γ : Ctx} → 𝒳 α Γ → 𝒴 α Γ
```

The concatenation of two contexts is denoted $\Gamma \mathbin{\dot{+}} \Delta$. An associated endofunctor on families is that of *context extension*: a $(\delta \mathbin{\Theta} \mathcal{X})$-term in context $\Gamma$ is an $\mathcal{X}$-term in the extended context $\Theta \mathbin{\dot{+}} \Gamma$.

$$
\begin{aligned}
&\_\mathbin{\dot{+}}\_ : \mathsf{Ctx} \to \mathsf{Ctx} \to \mathsf{Ctx} \\
&\emptyset \mathbin{\dot{+}} \Delta = \Delta \\
&(\alpha \cdot \Gamma) \mathbin{\dot{+}} \Delta = \alpha \cdot (\Gamma \mathbin{\dot{+}} \Delta)
\end{aligned}
\qquad
\begin{aligned}
&\delta : \mathsf{Ctx} \to \mathsf{Family}_s \to \mathsf{Family}_s \\
&\delta \mathbin{\Theta} \mathcal{X} \, \alpha \, \Gamma = \mathcal{X} \, \alpha \, (\Theta \mathbin{\dot{+}} \Gamma)
\end{aligned}
$$

*2.1.1 Cocartesian Structure of Contexts.* Fundamental to the algebraic study of abstract syntax is the *category of contexts* $\mathbb{F}$, with contexts $\Gamma, \Delta$ as objects and type-preserving mappings between variables of two contexts as morphisms. We will call these morphisms *renamings*; intuitively, a renaming $\rho : \Gamma \rightsquigarrow \Delta$ assigns a variable $\rho(v) : \alpha$ in context $\Delta$ to each variable $v : \alpha$ in $\Gamma$. Renamings can be generalised to arbitrary *context maps* $\sigma : \Gamma -[\, \mathcal{X} \,] \to \Delta$ that assign an $\mathcal{X}$-term $\sigma(v) : \mathcal{X} \, \alpha \, \Delta$ to each variable $v : \alpha$ in $\Gamma$ – a renaming is then expressed as an $\mathcal{I}$-valued context map.

$$
\begin{aligned}
&\_-[\_]\to\_ : \mathsf{Ctx} \to \mathsf{Family}_s \to \mathsf{Ctx} \to \mathsf{Set} \\
&\Gamma -[\, \mathcal{X} \,]\to \Delta = \{\alpha : T\} \to \mathcal{I} \, \alpha \, \Gamma \to \mathcal{X} \, \alpha \, \Delta
\end{aligned}
\qquad
\begin{aligned}
&\_\rightsquigarrow\_ : \mathsf{Ctx} \to \mathsf{Ctx} \to \mathsf{Set} \\
&\Gamma \rightsquigarrow \Delta = \Gamma -[\, \mathcal{I} \,]\to \Delta
\end{aligned}
$$

The category of contexts $\mathbb{F}$ is *cocartesian*, with context concatenation $\mathbin{\dot{+}}$ serving as coproduct. The injection maps $\Gamma \rightsquigarrow (\Gamma \mathbin{\dot{+}} \Delta)$ and $\Delta \rightsquigarrow (\Gamma \mathbin{\dot{+}} \Delta)$ and the universal copairing (definable for arbitrary $\mathcal{X}$-valued context maps) $\mathsf{copair} : (\Gamma -[\, \mathcal{X} \,]\to \Theta) \to (\Delta -[\, \mathcal{X} \,]\to \Theta) \to (\Gamma \mathbin{\dot{+}} \Delta) -[\, \mathcal{X} \,]\to \Theta$ can be constructed by pattern-matching on the contexts and variables, and repeated applications of old. A useful special case of copairing is adding a single term to a substitution, which corresponds to the standard *cons* operation in the theory of explicit substitutions [Abadi et al. 1989].

$$
\begin{aligned}
&\mathsf{add} : (\mathcal{X} : \mathsf{Family}_s) \to \mathcal{X} \, \alpha \, \Delta \to \Gamma -[\, \mathcal{X} \,]\to \Delta \to (\alpha \cdot \Gamma) -[\, \mathcal{X} \,]\to \Delta \\
&\mathsf{add} \, \mathcal{X} \, t \, \sigma = \mathsf{copair} \, \mathcal{X} \, (\lambda\{\, \mathsf{new} \to t \,\}) \, \sigma
\end{aligned}
$$

*2.1.2 Bicartesian Closed and Linear Structures of Families.* Since families are indexed sets, they inherit the bicartesian closed structure of **Set**: we take sums and products of families via pointwise disjoint unions and Cartesian products, and also define family exponentials pointwise as $(\mathcal{X} \Rightarrow \mathcal{Y}) \, \alpha \, \Gamma \triangleq \mathcal{X} \, \alpha \, \Gamma \to \mathcal{Y} \, \alpha \, \Gamma$ (since there are no functoriality requirements, relevant in the case of presheaf exponentials, to be satisfied). We also have the following *linear exponential* (derived from the Day [1970] internal hom) which will play an important role in metasubstitution:

$$
\begin{aligned}
&\_\multimap\_ : \mathsf{Family}_s \to \mathsf{Family}_s \to \mathsf{Family}_s \\
&(\mathcal{X} \multimap \mathcal{Y}) \, \alpha \, \Gamma = \{\Delta : \mathsf{Ctx}\} \to \mathcal{X} \, \alpha \, \Delta \to \mathcal{Y} \, \alpha \, (\Delta \mathbin{\dot{+}} \Gamma)
\end{aligned}
$$

## 2.2 Renaming and Substitution

Our next step is to precisely identify the structure required on a sorted family to support substitution; this way, defining substitution for a particular syntax will amount to equipping the family of terms of the syntax with such structure. Since, in practice, capture-avoiding substitution into syntactic terms involves weakening (a form of variable renaming), we are also looking for a way to axiomatise renaming as a basic fundamental notion. Our guiding principle throughout the development will be to characterise renaming, substitution, and other metatheoretic operations – along with their correctness laws – as natural, recognisable categorical constructions in sorted families.

*2.2.1 Renaming Structure.* A sorted family $\mathcal{X}$ admits renaming when $\mathcal{X}$-terms in one variable context can be compatibly transformed to $\mathcal{X}$-terms in another. The corresponding renaming operation lifts a renaming map $\Gamma \rightsquigarrow \Delta$ to a function $\mathcal{X} \, \alpha \, \Gamma \to \mathcal{X} \, \alpha \, \Delta$. This amounts to the requirement that the family $\mathcal{X} \, \alpha$ be a (covariant) *presheaf* on $\mathbb{F}$ (i.e. a functor from $\mathbb{F}$ to **Set**), and imposing such structure on every family would give rise to the presheaf model of Fiore et al. [1999]. Here, instead of going down this route, we refactor the type of the renaming operation on families as follows:

$$\mathsf{r} : \{\alpha : \mathsf{T}\}\{\Gamma : \mathsf{Ctx}\} \to \mathcal{X}\ \alpha\ \Gamma \to \big(\{\Delta : \mathsf{Ctx}\} \to (\Gamma \rightsquigarrow \Delta) \to \mathcal{X}\ \alpha\ \Delta\big)$$

Note that the codomain of $\mathsf{r}$ can be expressed as a function of $\alpha$ and $\Gamma$. By abstracting the renaming-dependence as a *modal operator* $\Box$ [Allais et al. 2021], we may then express the renaming operation as a map of families $\mathcal{X} \rightarrow \Box\,\mathcal{X}$, internalising the functorial action as a **Fam**$_\mathsf{s}$-morphism.

$$\Box : \mathsf{Family}_\mathsf{s} \to \mathsf{Family}_\mathsf{s}$$
$$\Box\,\mathcal{X}\ \alpha\ \Gamma = \{\Delta : \mathsf{Ctx}\} \to (\Gamma \rightsquigarrow \Delta) \to \mathcal{X}\ \alpha\ \Delta$$

The $\Box$ modality is a comonad and a coalgebra for it has to respect the identity and composition of renamings. This leads to our first observation: a sorted family has renaming structure when equipped with a $\Box$-*coalgebra* structure. The Coalg record collects these requirements.

```
record Coalg (𝒳 : Familyₛ) : Set where
    field   r      : 𝒳 ⇒ □ 𝒳
            counit : {t : 𝒳 α Γ} → r t id ≡ t
            comult : {ρ : Γ ⇝ Δ}{ϱ : Δ ⇝ Θ}{t : 𝒳 α Γ} → r t (ϱ ∘ ρ) ≡ r (r t ρ) ϱ
```

Since context transformations such as weakening, contraction, etc. correspond to renaming maps, the associated structural rules for a $\Box$-coalgebra $\mathcal{X}$ can be derived via renaming:

$$\mathsf{wkl} : \mathcal{X}\ \alpha\ \Gamma \to \mathcal{X}\ \alpha\ (\Gamma \mathbin{+} \Delta) \qquad \mathsf{wkr} : \mathcal{X}\ \alpha\ \Delta \to \mathcal{X}\ \alpha\ (\Gamma \mathbin{+} \Delta) \qquad \mathsf{contr} : \mathcal{X}\ \alpha\ (\Gamma \mathbin{+} \Gamma) \to \mathcal{X}\ \alpha\ \Gamma$$
$$\mathsf{wkl}\ t = \mathsf{r}\ t\ (\mathsf{inl}\ \Delta) \qquad\qquad\quad \mathsf{wkr}\ t = \mathsf{r}\ t\ (\mathsf{inr}\ \Gamma) \qquad\qquad\quad \mathsf{contr}\ t = \mathsf{r}\ t\ (\mathsf{copair}\ \mathcal{I}\ \mathsf{id}\ \mathsf{id})$$

The natural notion of a transformation between $\Box$-coalgebras is a *homomorphism*: a map $\mathcal{X} \rightarrow \mathcal{Y}$ that preserves the coalgebra structures of $\mathcal{X}$ and $\mathcal{Y}$.

```
record Coalg⇒ (𝒳□ : Coalg 𝒳)(𝒴□ : Coalg 𝒴) (f : 𝒳 ⇒ 𝒴) : Set where
    field   ⟨r⟩ : {ρ : Γ ⇝ Δ}{t : 𝒳 α Γ} → f (𝒳.r t ρ) ≡ 𝒴.r (f t) ρ
```

### 2.2.2 Pointed Structure.
If $\mathcal{X}$ is a sorted family of terms for a second-order abstract syntax, there must be a way to coerce variables into $\mathcal{X}$-terms. Sorted families with such a coercion $\eta : \mathcal{I} \rightarrow \mathcal{X}$ will be called *pointed*. If the underlying family of a $\Box$-coalgebra is pointed, we may also impose the requirement that the point is compatible with renaming. We characterise pointed coalgebras and their point-preserving homomorphisms as records:

```
record Coalg∗ (𝒳 : Familyₛ) : Set where       record Coalg∗⇒ (𝒳∗□ : Coalg∗ 𝒳) (𝒴∗□ : Coalg∗ 𝒴)
    field   □ : Coalg 𝒳  ;  η : 𝓘 ⇒ 𝒳                             (f : 𝒳 ⇒ 𝒴) : Set where
            roη :{v : 𝓘 α Γ}{ρ : Γ ⇝ Δ} →          field   □⇒ : Coalg⇒ 𝒳.□ 𝒴.□ f
                 r (η v) ρ ≡ η (ρ v)                             ⟨η⟩ : {v : 𝓘 α Γ} → f (𝒳.η v) ≡ 𝒴.η v
```

An example of a pointed $\Box$-coalgebra is the family of variables, with renaming implemented as application. The point $\eta$ of a pointed $\Box$-coalgebra is itself a pointed $\Box$-coalgebra homomorphism.

### 2.2.3 Substitution Structure.
The $\Box$ modality parametrises a family by a renaming. We now generalise this by parametrising a family $\mathcal{Y}$ by an arbitrary $\mathcal{X}$-valued context map. This is captured as a binary operation on families which we call the *internal substitution hom* of $\mathcal{X}$ and $\mathcal{Y}$:

$$(\!(\_,\_)\!) : \mathsf{Family}_\mathsf{s} \to \mathsf{Family}_\mathsf{s} \to \mathsf{Family}_\mathsf{s}$$
$$(\!(\ \mathcal{X}\ ,\ \mathcal{Y}\ )\!)\ \alpha\ \Gamma = \{\Delta : \mathsf{Ctx}\} \to (\Gamma \mathbin{-\!\lbrack} \mathcal{X}\ \rbrack\!\!\rightarrow \Delta) \to \mathcal{Y}\ \alpha\ \Delta$$

A renaming operation has, equivalently, the type $\mathcal{X} \rightarrow (\!(\mathcal{I}, \mathcal{X})\!)$, and a substitution operation on $\mathcal{X}$ has the type $\mathcal{X} \rightarrow (\!(\mathcal{X}, \mathcal{X})\!)$. The internal hom has a left adjoint $\odot$, called the *substitution tensor product*; that is, maps of the form $\mathcal{X} \rightarrow (\!(\mathcal{Y}, \mathcal{Z})\!)$ are naturally isomorphic to maps of the form $\mathcal{X} \odot \mathcal{Y} \rightarrow \mathcal{Z}$ via "uncurrying". These operators equip **Fam**$_\mathsf{s}$ with a skew-monoidal closed structure [Street 2013] (full monoidality would require that e.g. the unitor $\mathcal{I} \odot \mathcal{X} \to \mathcal{X}$ is invertible, which is not the case since $\odot$ is defined as a dependent sum, rather than a coend [Fiore et al. 1999]).

$$\_\odot\_ : \mathsf{Family_s} \to \mathsf{Family_s} \to \mathsf{Family_s}$$
$$(\mathcal{X} \odot \mathcal{Y})\, \alpha\, \Delta = \Sigma[\, \Gamma \in \mathsf{Ctx}\, ]\, (\mathcal{X}\, \alpha\, \Gamma \times (\Gamma -[\, \mathcal{Y}\, ]\!\to \Delta))$$

Expressed using the tensor product, the substitution operation has the type $\mathcal{X} \odot \mathcal{X} \to \mathcal{X}$ – it combines a term $\mathcal{X}\, \alpha\, \Gamma$ and substitution map $\Gamma -[\, \mathcal{X}\, ]\!\to \Delta$ into a term $\mathcal{X}\, \alpha\, \Delta$. To be proper, it must also be associative, and be compatible with the point of $\mathcal{X}$ when it has one. If we package this structure on a family $\mathcal{X}$ into a record – a point and substitution operation $\mathcal{I} \xrightarrow{\eta} \mathcal{X} \xleftarrow{\mu} \mathcal{X} \odot \mathcal{X}$ satisfying unit and associativity laws – we end up with precisely a *monoid* in $(\mathbf{Fam_s}, \mathcal{I}, \odot)$. Monoids can be equivalently expressed using the internal hom as $\mathcal{I} \xrightarrow{\eta} \mathcal{X} \to (\!|\, \mathcal{X}, \mathcal{X}\, |\!)$, which is the presentation we prefer for technical reasons: the metatheory proofs would not go through if we had used $\odot$.

```
record Mon (M : Familyₛ) : Set where
  field η : I ⇀ M
        μ : M ⇀ (| M , M |)

        lunit : {σ : Γ −[ M ]→ Δ}{v : I α Γ} → μ (η v) σ ≡ σ v
        runit : {t : M α Γ} → μ t η ≡ t
        assoc : {σ : Γ −[ M ]→ Δ} {ς : Δ −[ M ]→ Θ} {t : M α Γ} →
                μ (μ t σ) ς ≡ μ t (λ v → μ (σ v) ς)
```

The multiplication $\mu$ represents simultaneous substitution, replacing every variable in context $\Gamma$ with an $\mathcal{M}$-term in $\Delta$. In practice (e.g. in $\beta$-reduction) one often uses one- or two-variable substitution for the last variable or variables in the context, which is derived using add:

$$[\_/] : M\, \alpha\, \Gamma \to M\, \beta\, (\alpha \cdot \Gamma) \to M\, \beta\, \Gamma \qquad [\_,\_/]_2 : M\, \alpha\, \Gamma \to M\, \beta\, \Gamma \to M\, \tau\, (\alpha \cdot \beta \cdot \Gamma) \to M\, \tau\, \Gamma$$
$$[\, s\, /]\, t = \mu\, t\, (\text{add}\, M\, s\, \eta) \qquad\qquad [\, s_1, s_2\, /]_2\, t = \mu\, t\, (\text{add}\, M\, s_1\, (\text{add}\, M\, s_2\, \eta))$$

Monoid homomorphisms preserve the unit and multiplication. When $\mathcal{M}$ is a family associated with an inductively defined syntax, $\mathcal{N}$ is a model of the syntax, and $f$ is a map $\mathcal{M} \to \mathcal{N}$, the preservation of multiplication $\langle \mu \rangle : \{\sigma : \Gamma -[\, M\, ]\!\to \Delta\}\{t : M\, \alpha\, \Gamma\} \to f\, (M.\mu\, t\, \sigma) \equiv \mathcal{N}.\mu\, (f\, t)\, (f \circ \sigma)$ expresses the *semantic substitution lemma*: the interpretation of substitution in the syntax is the substitution of interpretations in the model. We give its familiar form for single-variable substitution as an example below. The fact that $f$ commutes with add is established by function extensionality, case analysis on the variable, and preservation of the unit $\langle \eta \rangle : \{v : \mathcal{I}\, \alpha\, \Gamma\} \to f\, (M.\eta\, v) \equiv \mathcal{N}.\eta\, v$.

```
sub-lemma : (s : M α Γ)(t : M β (α · Γ)) → f (M.[ s /] t) ≡ N.[ f s /] (f t)
sub-lemma s t = trans ⟨μ⟩ (cong (N.μ (f t)) (ext λ{ new → refl ; (old y) → ⟨η⟩}))
```

Every monoid has an induced pointed $\square$-coalgebra instance $\mathcal{M}_*^{\square} : \mathsf{Coalg_*}$, where renaming is implemented by substituting variables for variables, and pointed $\square$-coalgebra laws follow from monoid axioms. But, what if $\mathcal{M}$ already comes with a coalgebra structure? We will revisit this question after examining some general properties of maps into internal homs.

## 2.3 Parametrised Maps

Working with families with coalgebra structure lets us be precise about when renaming is required in constructions: for example, we need the coalgebra structure for weakening on $\mathcal{X}$, but not for copairing of $\mathcal{X}$-valued context maps. However, as Fiore et al. [1999] show, the interaction between the substitution tensor and the presheaf structure forms a core part of the model theory that will need to be recast in our coalgebraic setting. To achieve this, we introduce an important collection of properties for maps into internal homs (equivalently, out of substitution tensors).

*2.3.1 Coalgebraic Maps.* Maps of the form $f : \mathcal{X} \twoheadrightarrow (\!(\,\mathcal{P}\,,\,\mathcal{Y}\,)\!)$ play a vital role in the mathematical theory, since they transform $\mathcal{X}$ to $\mathcal{Y}$ while altering the variable context according to a parameter family $\mathcal{P}$. Many syntactic operations are of this form: for example, as we have already mentioned, renaming $\mathcal{X} \twoheadrightarrow (\!(\,\mathcal{I}\,,\,\mathcal{X}\,)\!)$ and substitution $\mathcal{X} \twoheadrightarrow (\!(\,\mathcal{X}\,,\,\mathcal{X}\,)\!)$ are maps parametrised by $\mathcal{I}$ and $\mathcal{X}$, respectively. If all three families $\mathcal{X}, \mathcal{P}, \mathcal{Y}$ have pointed coalgebra structure, there are two ways in which a map $f$ can be compatible with the various renaming operations: it can preserve renaming, or identify terms that get renamed via the coalgebra structure and via the context map. Furthermore, one can state a compatibility law between the points of all three families. Maps which satisfy these three properties are called *coalgebraic*.

<pre>
record Coalgebraic (f : 𝒳 ⇸ (( 𝒫 , 𝒴 ))) : Set where
    field  rof : {σ : Γ −[ 𝒫 ]→ Δ}{ϱ : Δ ⤳ Θ} {t : 𝒳 α Γ} → 𝒴.r (f t σ) ϱ  ≡ f t (λ v → 𝒫.r (σ v) ϱ)
           for : {ρ : Γ ⤳ Δ}{ς : Δ −[ 𝒫 ]→ Θ} {t : 𝒳 α Γ} → f (𝒳.r t ρ) ς  ≡ f t (ς ∘ ρ)
           foη :                                  {v : 𝓘 α Γ} → f (𝒳.η v) 𝒫.η ≡ 𝒴.η v
</pre>

Though the hom of pointed families is not in general pointed, the codomain of coalgebraic maps has a pointed □-coalgebra structure, and the map itself is a pointed □-coalgebra homomorphism.

<pre>
Cod*□ : Coalg* (( 𝒫 , 𝒴 ))
Cod*□ = record { η = λ v σ → f (𝒳.η v) σ
                ; □ = record { r = λ h ρ σ → h (σ ∘ ρ) ; counit = refl ; comult = refl }
                ; rσ = λ {v}{ρ} → ext λ σ → begin f (𝒳.η v) (σ ∘ ρ)   ≡⟨ for ⟩
                                                    f (𝒳.r (𝒳.η v) ρ) σ ≡⟨ 𝒳.rση ⟩
                                                    f (𝒳.η (ρ v)) σ     ∎ }

f*□⇒ : Coalg*⇒ 𝒳*□ Cod*□ f
f*□⇒ = record { □⇒ = record { ⟨r⟩ = ext (λ σ → foη) } ; ⟨η⟩ = refl }
</pre>

Examples of coalgebraic maps are the renaming map for a pointed □-coalgebra:

<pre>
rᶜ : (𝒳*□ : Coalg* 𝒳) → Coalgebraic 𝒳*□ 𝓘*□ 𝒳*□ 𝒳.r
rᶜ 𝒳*□ = record { rof = sym comult ; for = sym comult ; foη = rση }
</pre>

as well as the multiplication $\mu$ for a substitution monoid and the structural map $\mathsf{j} : \mathcal{I} \twoheadrightarrow (\!(\,\mathcal{X}\,,\,\mathcal{X}\,)\!)$ of skew-closed categories that corresponds to application (mapping $v$ to $\sigma \mapsto \sigma(v)$). Later on, we will be able to show that universal parametrised maps from an initial syntactic algebra are coalgebraic, which will be a prerequisite for proving the substitution axioms.

*2.3.2 Lifting and Strength.* The definition of simultaneous substitution by hand involves traversing the host term, applying the substitution map to variables, and recursing into subterms. The difficulties arise when the subterm has a newly bound variable, such as in the body of a $\lambda$-abstraction. To avoid variable shadowing, substitution must map the newly bound variable to itself; to avoid variable capture, the terms to be substituted must not contain free occurrences of the new variable. Fortunately, intrinsically-typed encoding guides the complex de Bruijn "arithmetic" required and guarantees to maintain type- and scope-safety.

The critical step is applying a substitution $\sigma : \Gamma -[ \mathcal{P} ]\to \Delta$ to a term in an extended context $\mathcal{P}\ \beta\ (\alpha \cdot \Gamma)$, without disturbing the newly bound variable $\alpha$. The usual name for the transformation needed on $\sigma$ is *lifting* [Altenkirch and Reus 1999; Benton et al. 2012], which can be generalised to arbitrary extensions as $\mathsf{lift}\ \Theta\ \sigma : (\Theta + \Gamma) -[ \mathcal{P} ]\to (\Theta + \Delta)$. The inductive definition of lifting requires both a point (to map newly bound variables to themselves) and renaming (to weaken the context of the recursive call), so we demand the structure of a pointed $\square$-coalgebra on $\mathcal{P}$:

$$\text{lift} : (\Theta : \text{Ctx}) \to (\Gamma -[\, \mathcal{P} \,] \to \Delta) \to (\Theta \mathbin{\dot{+}} \Gamma) -[\, \mathcal{P} \,] \to (\Theta \mathbin{\dot{+}} \Delta)$$
$$\text{lift } \emptyset \quad\quad \sigma\, v \quad\quad = \sigma\, v$$
$$\text{lift } (\tau \cdot \Theta)\, \sigma \text{ new} \quad = \mathcal{P}.\eta \text{ new}$$
$$\text{lift } (\tau \cdot \Theta)\, \sigma \text{ (old } v) = \mathcal{P}.\text{r (lift } \Theta\, \sigma\, v) \text{ old}$$

While this definition works, it expresses lift as a Set-level transformation of context maps, rather than as a morphism of sorted families – as such, it counterposes our goal of representing metasyntactic operations purely algebraically. Consequently, it is not a priori obvious what laws lift should satisfy. Benton et al. [2012, Section 4] demonstrate the intricate inter-dependence of lift with the other structures by listing eight laws that concern the interaction amongst lifting, renaming, and substitution, all of which must be proved in order, with each property building on top of the previous ones. It is these sorts of auxiliary definitions and ad-hoc lemmas that we wish to avoid in our systematic, categorical development of abstract syntax.

The key to demystifying lift lies in the categorical concept of *cotensorial strength* [Kock 1971] for a sorted-family endofunctor $F$: namely, an operation of type $F\llbracket\, \mathcal{X}, \mathcal{Y} \,\rrbracket \to \llbracket\, \mathcal{X}, F\mathcal{Y} \,\rrbracket$ satisfying certain coherence laws. The analogous tensorial strength $F(\mathcal{X}) \odot \mathcal{Y} \to F(\mathcal{X} \odot \mathcal{Y})$ plays a central role in the presheaf model of abstract syntax [Fiore 2008], since it captures the intuition of pushing a substitution into a syntactic structure represented by the endofunctor $F$. Most notably, lift attains a natural place within the categorical model as it can be used to implement strength for the instance $F = \delta\,\Theta$: the resulting operation $\delta\,\Theta \llbracket\, \mathcal{P}, \mathcal{X} \,\rrbracket \to \llbracket\, \mathcal{P}, \delta\,\Theta\mathcal{X} \,\rrbracket$ is then responsible for pushing a substitution $\sigma\colon \Gamma -[\, \mathcal{P} \,] \to \Delta$ under a binder of variables in context $\Theta$.

In our setting of families and coalgebras, a more refined notion of strength is required (similar to the structural strength of Borthelle et al. [2020]). A *coalgebraic strength* for $F$ is a transformation

$$\text{str} : F\llbracket\, \mathcal{P}, \mathcal{X} \,\rrbracket \to \llbracket\, \mathcal{P}, F\mathcal{X} \,\rrbracket$$

where $\mathcal{P}$ is a pointed $\square$-coalgebra and $\mathcal{X}$ is a sorted family. The operation must be *natural* in both components: that is, it must suitably commute with the functorial mapping of any pointed $\square$-coalgebra homomorphism $f\colon \mathcal{Q} \to \mathcal{P}$ and family of maps $g\colon \mathcal{X} \to \mathcal{Y}$.

$$\text{str-nat}_1 : (f_*^\square \Rightarrow : \text{Coalg}_* \Rightarrow \mathcal{Q}_*^\square\, \mathcal{P}_*^\square\, f)\, (h : F\llbracket\, \mathcal{P}, \mathcal{X} \,\rrbracket\, \alpha\, \Gamma)\, (\sigma : \Gamma -[\, \mathcal{Q} \,] \to \Delta) \to$$
$$\text{str } \mathcal{P}_*^\square\, \mathcal{X}\, h\, (f \circ \sigma) \equiv \text{str } \mathcal{Q}_*^\square\, \mathcal{X}\, (F_1\, (\lambda\, h'\, \sigma' \to h'\, (f \circ \sigma'))\, h)\, \sigma$$

$$\text{str-nat}_2 : (g : \mathcal{X} \to \mathcal{Y})(h : F\llbracket\, \mathcal{P}, \mathcal{X} \,\rrbracket\, \alpha\, \Gamma)(\sigma : \Gamma -[\, \mathcal{P} \,] \to \Delta) \to$$
$$\text{str } \mathcal{P}_*^\square\, \mathcal{Y}\, (F_1\, (\lambda\, h'\, \sigma' \to g\, (h'\, \sigma'))\, h)\, \sigma \equiv F_1\, g\, (\text{str } \mathcal{P}_*^\square\, \mathcal{X}\, h\, \sigma)$$

The strength also satisfies a unit law str-unit: $(h : F\llbracket\, \mathcal{I}, \mathcal{X} \,\rrbracket\, \alpha\, \Gamma) \to \text{str } \mathcal{I}_*^\square\, \mathcal{X}\, h\, \text{id} \equiv F_1\, (\text{i } \mathcal{X})\, h$ and an associativity law with respect to the two other structural transformations of skew-closed categories: the unit $\text{i} = \lambda\, h \to h\, \text{id} : \llbracket\, \mathcal{I}, \mathcal{X} \,\rrbracket \to \mathcal{X}$ and the (curried) composition of internal homs $\text{L} = \lambda\, h\, \sigma\, \varsigma \to h\, (\lambda\, v \to \sigma\, v\, \varsigma) : \llbracket\, \mathcal{Y}, \mathcal{Z} \,\rrbracket \to \llbracket\, \llbracket\, \mathcal{X}, \mathcal{Y} \,\rrbracket, \llbracket\, \mathcal{X}, \mathcal{Z} \,\rrbracket \,\rrbracket$. The associativity law has to be stated in terms of coalgebraic maps $f : \mathcal{P} \to \llbracket\, \mathcal{Q}, \mathcal{R} \,\rrbracket$, since the usual pentagon identity [Kock 1971, Lemma 1.3] is too strict in the skew-closed setting (see str-assoc in the diagram below). This generalised associativity law neatly combines with naturality to give the following powerful corollary, given both as Agda code and – for a clearer presentation – as a commutative diagram:

$$\mathsf{str\text{-}dist} : \{f : \mathcal{P} \rightarrowtail \llparenthesis\, Q\,,\mathcal{R}\,\rrparenthesis\} \, (f^{c} : \mathsf{Coalgebraic}\ \mathcal{P}^{\square}_{*}\ Q^{\square}_{*}\ \mathcal{R}^{\square}_{*}\ f)$$
$$(h : F \llparenthesis\,\mathcal{R}\,,\mathcal{X}\,\rrparenthesis\ \alpha\ \Gamma)(\sigma : \Gamma -\![\ \mathcal{P}\ ]\!\rightarrow \Delta)(\varsigma : \Delta -\![\ Q\ ]\!\rightarrow \Theta) \rightarrow$$
$$\mathsf{str}\ \mathcal{R}^{\square}_{*}\ \mathcal{X}\ h\ (\lambda\ v \rightarrow f\ (\sigma\ v)\ \varsigma)$$
$$\equiv\ \mathsf{str}\ Q^{\square}_{*}\ \mathcal{X}\ (\mathsf{str}\ \mathcal{P}^{\square}_{*}\ \llparenthesis\, Q\,,\mathcal{X}\,\rrparenthesis\ (F_{1}\ (\lambda\ h\ \varsigma \rightarrow h\ (\lambda\ v \rightarrow f\ (\sigma\ v)\ \varsigma))\ h)\ \sigma)\ \varsigma$$

As $f$ is coalgebraic, its codomain $\llparenthesis\, Q\,,\mathcal{R}\,\rrparenthesis$ is a pointed $\square$-coalgebra (Section 2.3.1) and thus is a valid first component to $\mathsf{str}_{\llparenthesis Q,\mathcal{R}\rrparenthesis,\llparenthesis Q,\mathcal{X}\rrparenthesis}$. Remarkably, for different choices of $f$, the str-dist corollary above generalises all four lifting lemmas given by Benton et al., fulfilling our goal of placing lift and its laws on a formal foundation.

The axiomatisation of strength as a categorical concept is required for the initiality and freeness theorems of Section 3. The strength for a signature endofunctor (Section 4.2) will invariably derive from the Strength instance of the context extension endofunctor $\delta$, whose implementation makes use of lift: to feed a context map $\Gamma -\![\ \mathcal{P}\ ]\!\rightarrow \Delta$ into a hom $\llparenthesis\,\mathcal{P}\,,\mathcal{X}\,\rrparenthesis\ \alpha\ (\Theta + \Gamma)$, one has to extend both its domain and codomain with $\Theta$. The strength proofs feature many of the intricate properties we axiomatised earlier, such as homomorphism and coalgebraic laws. We refer the reader to the formalisation for the details.

*2.3.3   Coalgebraic Monoids.* We revisit the question posed at the end of Section 2.2.3. As shown there, every substitution monoid is a $\square$-coalgebra, since renaming with $\rho : \Gamma \rightsquigarrow \Delta$ can be implemented as substitution with $\eta \circ \rho : \Gamma -\![\ \mathcal{X}\ ]\!\rightarrow \Delta$. This may suggest that renaming for a syntax can be derived from substitution, and so that one only needs to define the latter operation. Attempting this will be futile, however: substitution into terms with variable binding will require weakening (as part of lift), a special case of renaming. Consequently, to equip a family of terms with substitution structure, one needs to have already shown that it is a pointed $\square$-coalgebra.

This a priori renaming structure will not necessarily be equivalent to the one induced by substitution (unlike in the presheaf model, where the two are identified by the quotiented dependent sum used in the definition of the substitution tensor product), but their equivalence is a prerequisite of the free monoid proof in Theorem 3.1 below. We overcome this conflict by axiomatising families with compatible pointed coalgebra and monoid structures, called *coalgebraic monoids*:

$$\mathsf{record\ CoalgMon}\ (\mathcal{X} : \mathsf{Family_S}) : \mathsf{Set\ where}$$
$$\mathsf{field}\quad \mathcal{X}^{\square}_{*} : \mathsf{Coalg_*}\ \mathcal{X}$$
$$\mathcal{X}^{\mathsf{M}} : \mathsf{Mon}\quad \mathcal{X}$$
$$\eta\text{-compat} : \qquad\qquad \{v : \mathcal{I}\ \alpha\ \Gamma\} \rightarrow \eta^{\square}_{*}\ v \equiv \eta^{\mathsf{M}}\ v$$
$$\mu\text{-compat} : \{\rho : \Gamma \rightsquigarrow \Delta\}\ \{t : \mathcal{X}\ \alpha\ \Gamma\} \rightarrow \mathsf{r}\ t\ \rho \equiv \mu\ t\ (\eta^{\mathsf{M}} \circ \rho)$$

The compatibility laws ensure that the existing $\mathcal{X}^{\square}_{*}$ and derived $\mathcal{X}^{\mathsf{M},\square}_{*}$ pointed $\square$-coalgebra structures on $\mathcal{X}$ are equivalent, and, in particular, can be exchanged in the first component of str.

We have now set up all the mathematical foundations needed for the development. Next, we move on to the central conceptual tool of our framework: initial algebra semantics.

## 3  INITIAL ALGEBRA SEMANTICS

Initial algebra semantics [Goguen et al. 1976] is one of the most useful concepts in the study of data types and functional programming. It stems from the observation that every inductive data type $\mathsf{T}$ corresponds to an endofunctor $F_T$ of which the data type is an initial algebra. For example, the data type $\mathbb{N}$ of natural numbers has associated endofunctor $F_{\mathbb{N}}(A) = 1 + A$ and comes equipped with an isomorphism $1 + \mathbb{N} \xrightarrow{\cong} \mathbb{N}$. Furthermore, for any $F_{\mathbb{N}}$-algebra $A$ – equivalently, a type $A$ with an element $z\colon A$ and an operation $s\colon A \to A$ – one has a unique function $\mathrm{rec}_{\mathbb{N}}(z,s)\colon \mathbb{N} \to A$ that satisfies $\mathrm{rec}_{\mathbb{N}}(z,s)(0) = z$ and $\mathrm{rec}_{\mathbb{N}}(z,s)(n+1) = s(\mathrm{rec}_{\mathbb{N}}(z,s)(n))$. This is usually called the *recursor*, *fold*, or *catamorphism* for $\mathbb{N}$ and captures the process of consuming an inhabitant of the inductive type by recursion on its structure.

This idea extends to inductively defined families of types and endofunctors thereon [Altenkirch et al. 2015], such as a data type of intrinsically-typed terms. In particular, the recursor out of an initial family of terms can then be seen as a compositional semantic interpretation map: for example, an interpretation of the STLC in any of its models. As we explore next, initial algebra semantics is not only useful for semantic purposes: it can also be used to implement syntactic operations and prove their laws for free.

### 3.1  Signature Algebras and Monoids

Just as the categorical theory, our framework is entirely signature-generic. The metatheory developed here can be freely instantiated for any second-order signature, encompassing algebraic theories, computational calculi, logics, etc. Thus, for the remainder of the section, we fix a sorted-family endofunctor $\Sigma\colon \mathbf{Fam}_{\mathsf{s}} \to \mathbf{Fam}_{\mathsf{s}}$ with an instance of coalgebraic strength $\Sigma{:}\mathrm{Str}$. Algebras $\Sigma\mathcal{A} \rightsquigarrow \mathcal{A}$ for this endofunctor represent families that support the operations of the signature. For example, for the signature $\Sigma_{\Lambda}$ of the STLC, a $\Sigma_{\Lambda}$-algebra is a family $\mathcal{A}$ equipped with operations app$\colon \mathcal{A}\,(\alpha \rightarrowtail \beta)\,\Gamma \times \mathcal{A}\,\alpha\,\Gamma \to \mathcal{A}\,\beta\,\Gamma$ and abs$\colon \mathcal{A}\,(\beta\cdot(\tau\cdot\Gamma)) \to \mathcal{A}\,(\alpha \rightarrowtail \beta)\,\Gamma$.

We also consider $\Sigma$-algebras with substitution structure, known as $\Sigma$-*monoids* [Fiore et al. 1999; Fiore and Saville 2017]. These are families with algebra structure $\Sigma\mathcal{X} \rightsquigarrow \mathcal{X}$ and monoid structure $\mathcal{I} \rightsquigarrow \mathcal{X} \rightsquigarrow (\!(\,\mathcal{X}, \mathcal{X}\,)\!)$ which are compatible with each other; the compatibility is expressed using the coalgebraic strength str on $\Sigma$, used to swap the constructor application with the parametrisation by a context map. Note that str uses the pointed $\square$-coalgebra structure derived from substitution.

record $\Sigma$Mon $(\mathcal{X}: \mathrm{Family}_{\mathsf{s}})$ : Set where

field $\mathcal{X}^{\mathrm{M}}$    : Mon $\mathcal{X}$

  alg    : $\Sigma\mathcal{X} \rightsquigarrow \mathcal{X}$

  $\mu\langle \mathrm{alg}\rangle$ : $\{\sigma : \Gamma -\!\!\lbrack\,\mathcal{X}\,\rbrack\!\!\to \Delta\}\{t : \Sigma\mathcal{X}\,\alpha\,\Gamma\} \to$
        $\mu\,(\mathrm{alg}\,t)\,\sigma \equiv \mathrm{alg}\,(\mathrm{str}\,\mathcal{X}^{\square}_{*}\,\mathcal{X}\,(\Sigma_1\,\mu\,t)\,\sigma)$

$$\Sigma\mathcal{X} \xrightarrow{\Sigma\mu} \Sigma(\!(\,\mathcal{X},\mathcal{X}\,)\!) \xrightarrow{\mathrm{str}_{\mathcal{X},\mathcal{X}}} (\!(\,\mathcal{X},\Sigma\mathcal{X}\,)\!)$$
$$\mathrm{alg}\big\downarrow \qquad\qquad\qquad\qquad \big\downarrow(\!(\,\mathrm{id},\mathrm{alg}\,)\!)$$
$$\mathcal{X} \xrightarrow{\hspace{3cm}\mu\hspace{3cm}} (\!(\,\mathcal{X},\mathcal{X}\,)\!)$$

### 3.2  Algebras with Metavariables

Our formalisation is novel in that it also incorporates *parametrised metavariables* [Aczel 1978; Hamana 2004; Fiore 2008]: terms that take the syntactic form $\mathfrak{m}\langle -_1, \cdots, -_{\ell}\rangle$ and are to be imagined as "named holes with slots", with the "hole" named $\mathfrak{m}$ standing for an unspecified term with $\ell$ open variables and the "slots" containing terms that occupy the open variables. They too come from a sorted family: for instance, the family $\mathfrak{B}$ containing the metavariables $\mathfrak{m}\colon \mathfrak{B}\,\beta\,[\alpha]$ and $\mathfrak{n}\colon \mathfrak{B}\,\alpha\,[\,]$ can be used to construct terms such as $r = \big(\lambda x : \alpha.\,\mathfrak{m}\langle x\rangle\big)\big(\mathfrak{n}\langle\rangle\big)$ and $t = \mathfrak{m}\langle \mathfrak{n}\langle\rangle\rangle$. Metavariables let us abstractly reason about generic terms of some specific syntactic structure: for example, the notion of $\beta$-equivalence can be axiomatised as the identification of all common instances of $r$ and $t$ above. For the rest of this section until Section 3.5, we also fix a sorted family $\mathfrak{X}$ of metavariables.

A $\Sigma$-algebra structure map on $\mathcal{A}$ captures all the constructors of a second-order syntax, but for full generality, one also needs to account for both variables and metavariables. These are respectively represented by a point $\mathsf{var} : \mathcal{I} \rightarrow \mathcal{A}$ and a metavariable operator $\mathsf{mvar} : \mathfrak{X} \rightarrow \llparenthesis \mathcal{A}, \mathcal{A} \rrparenthesis$. In elementary terms, the latter is a function that associates an $\mathcal{A}$-term $\mathsf{mvar}\ \mathfrak{m}\ \varepsilon : \mathcal{A}\, \tau\, \Gamma$ to every parametrised metavariable $\mathfrak{m} : \mathfrak{X}\, \tau\, \Pi$ and metavariable environment $\varepsilon : \Pi -\!\!\!\left[\, \mathcal{A}\,\right] \rightarrow \Gamma$. For example, the term $\mathfrak{m}\langle t, s \rangle : \mathcal{A}\, \tau\, \Gamma$ for a metavariable $\mathfrak{m} : \mathfrak{X}\, \tau\, [\alpha, \beta]$, and terms $t : \mathcal{A}\, \alpha\, \Gamma$ and $s : \mathcal{A}\, \beta\, \Gamma$ is represented by $\mathsf{mvar}\ \mathfrak{m}\ \lambda\{\ \mathsf{new} \rightarrow t\ ;\ \mathsf{old\ new} \rightarrow s\ \}$. Families that support this structure will be called $(\Sigma, \mathfrak{X})$-*meta-algebras*, with the expected notion of homomorphism between them.

$$\mathsf{record\ MetaAlg\ }(\mathcal{A} : \mathsf{Family_s}) : \mathsf{Set\ where}$$
$$\mathsf{field}\quad \mathsf{alg}\quad : \Sigma\, \mathcal{A} \rightarrow \mathcal{A}$$
$$\mathsf{var}\quad : \quad \mathcal{I} \rightarrow \mathcal{A}$$
$$\mathsf{mvar} : \quad \mathfrak{X} \rightarrow \llparenthesis \mathcal{A}, \mathcal{A} \rrparenthesis$$

$$\mathsf{record\ MetaAlg}{\Rightarrow}\ (\mathcal{A}^\Sigma : \mathsf{MetaAlg}\, \mathcal{A})\ (\mathcal{B}^\Sigma : \mathsf{MetaAlg}\, \mathcal{B})\ (f : \mathcal{A} \rightarrow \mathcal{B}) : \mathsf{Set\ where}$$
$$\mathsf{field}\quad \langle \mathsf{alg} \rangle\quad : \{t : \Sigma\, \mathcal{A}\, \alpha\, \Gamma\}\qquad\qquad\qquad \rightarrow f\, (\mathcal{A}.\mathsf{alg}\, t)\qquad \equiv \mathcal{B}.\mathsf{alg}\, (\Sigma\, f\, t)$$
$$\langle \mathsf{var} \rangle\quad : \{v : \mathcal{I}\, \alpha\, \Gamma\}\qquad\qquad\qquad \rightarrow f\, (\mathcal{A}.\mathsf{var}\, v)\qquad \equiv \mathcal{B}.\mathsf{var}\, v$$
$$\langle \mathsf{mvar} \rangle : \{\mathfrak{m} : \mathfrak{X}\, \alpha\, \Pi\}\{\varepsilon : \Pi -\!\!\!\left[\, \mathcal{A}\,\right] \rightarrow \Gamma\} \rightarrow f\, (\mathcal{A}.\mathsf{mvar}\, \mathfrak{m}\, \varepsilon) \equiv \mathcal{B}.\mathsf{mvar}\, \mathfrak{m}\, (f \circ \varepsilon)$$

Such meta-algebras and their homomorphisms form a category **MetaAlg**, whose initial object – whenever it exists – will be denoted $\mathbb{T}\, \mathfrak{X}$ (or just $\mathbb{T}$, if the metavariable family is clear from the context) with structural maps $\mathsf{alg}$, $\mathsf{var}$ and $\mathsf{mvar}$. The universal property of initial objects states that there is a unique meta-algebra homomorphism $\mathsf{sem} : \mathbb{T} \rightarrow \mathcal{A}$ into any meta-algebra $\mathcal{A}$. Note that, varying $\mathfrak{X}$, $\mathbb{T}$ acts as a mapping from families $\mathfrak{X}$ to $(\Sigma, \mathfrak{X})$-meta-algebras $\mathbb{T}\, \mathfrak{X}$. Adapted from Fiore [2008, Theorem 2], we have the following main result in the study of abstract syntax:

**Theorem 3.1.** *The initial meta-algebra $\mathbb{T}\, \mathfrak{X}$ is the free $\Sigma$-monoid on $\mathfrak{X}$.*

The most interesting implication of this concise statement is that the substitution structure of the free $\Sigma$-monoid is induced purely by initiality. Since the initial meta-algebra $\mathbb{T}$ corresponds to the inductively defined grammar of terms (see Section 4.3), it follows that the syntactic structure fully determines the action of substitution. The theorem formally captures the observation that much of syntactic metatheory is uninteresting boilerplate, and the methodology of initial algebra semantics will allow us to extract syntactic, compositional interpretations, and related correctness laws from the syntax for free. We outline the proof of the theorem in the rest of the section.

### 3.3 Parametrised Interpretations

The initial algebra approach explains and validates our adherence to formality and efforts to internalise syntactic operations as categorical constructions at the level of sorted families. Since they are maps out of an initial meta-algebra, the renaming $\mathbb{T} \rightarrow \Box\mathbb{T}$ and the substitution $\mathbb{T} \rightarrow \llparenthesis \mathbb{T}, \mathbb{T} \rrparenthesis$ operations may be induced by initiality as soon as we show that their codomains $\Box\mathbb{T}$ and $\llparenthesis \mathbb{T}, \mathbb{T} \rrparenthesis$ are meta-algebras. This will be derived from the following lemma.

**Lemma 3.2.** *Given a pointed $\Box$-coalgebra $\mathcal{P}$, a $\Sigma$-algebra $\mathcal{A}$, and family maps $\varphi : \mathcal{P} \rightarrow \mathcal{A}$ and $\chi : \mathfrak{X} \rightarrow \llparenthesis \mathcal{A}, \mathcal{A} \rrparenthesis$, the internal hom $\llparenthesis \mathcal{P}, \mathcal{A} \rrparenthesis$ acquires a $(\Sigma, \mathfrak{X})$-meta-algebra structure.*

The proof is encapsulated in the Traversal module, instantiations of which will give rise to homomorphic initial algebra interpretations $\mathbb{T} \rightarrow \llparenthesis \mathcal{P}, \mathcal{A} \rrparenthesis$. It crucially relies on the coalgebraic strength when defining the structure map $\Sigma \llparenthesis \mathcal{P}, \mathcal{A} \rrparenthesis \rightarrow \llparenthesis \mathcal{P}, \Sigma\mathcal{A} \rrparenthesis \rightarrow \llparenthesis \mathcal{P}, \mathcal{A} \rrparenthesis$. A simple corollary (derived by instantiating $\mathcal{P}$ with $\mathcal{I}$) is that if $\mathcal{A}$ is a meta-algebra, then so is $\Box\, \mathcal{A}$.

```
module Traversal (𝒫*□ : Coalg* 𝒫) (alg𝒜 : Σ 𝒜 ⇀ 𝒜) (φ : 𝒫 ⇀ 𝒜) (χ : 𝔛 ⇀ ⦅ 𝒜 , 𝒜 ⦆) where
  TravΣ : MetaAlg ⦅ 𝒫 , 𝒜 ⦆
  TravΣ = record { alg   = λ h    σ → alg𝒜 (str 𝒫*□ 𝒜 h σ)
                 ; var   = λ v    σ → φ (σ v)
                 ; mvar  = λ 𝔪 ε  σ → χ 𝔪 (λ v → ε v σ) }
```

One may be tempted to immediately induce the substitution map $\mathbb{T} \rightharpoonup ⦅ \mathbb{T} , \mathbb{T} ⦆$ as a $\mathbb{T}$-parametrised traversal into $\mathbb{T}$. However, that would require a pointed □-coalgebra instance on $\mathbb{T}$, which we do not yet have. Our formalism very concretely exhibits the dependence of substitution on renaming.

### 3.4 Σ-Monoid Structure by Initiality

The construction of the renaming and substitution maps on the initial meta-algebra $\mathbb{T}$ make extensive use of *initiality*: maps of the form $\mathbb{T} \rightharpoonup 𝒜$ are uniquely induced by equipping $𝒜$ with a meta-algebra structure. The proofs of the renaming and substitution laws are then established by proving that the (composite) maps that correspond to the two sides of an equation are meta-algebra homomorphisms and must therefore be equal.

As an example, the renaming map $\mathtt{ren} : \mathbb{T} \rightharpoonup □\mathbb{T}$ is induced as the unique homomorphism from $\mathbb{T} \in \mathbf{MetaAlg}$ to the sorted family $□\mathbb{T}$ regarded as a meta-algebra by means of Lemma 3.2. The counit law $\mathtt{ren}\ t\ \mathrm{id} = t$ amounts to showing that the mapping $t \mapsto \mathtt{ren}\ t\ \mathrm{id} : \mathbb{T} \rightharpoonup \mathbb{T}$ is a meta-algebra homomorphism so that – by initiality of $\mathbb{T}$ – must be equal to the identity $t \mapsto t$. Similarly, proving that the **Fam**ₛ-morphisms defined as $t, \rho, \varrho \mapsto \mathtt{r}\ t\ (\varrho \circ \rho)$ and $t, \rho, \varrho \mapsto \mathtt{r}\ (\mathtt{r}\ t\ \rho)\ \varrho$ of type $\mathbb{T} \rightharpoonup ⦅ \mathcal{I} , ⦅ \mathcal{I} , \mathbb{T} ⦆ ⦆$ are meta-algebra homomorphisms implies that they must be equal, giving us the comultiplication law. The proofs (available in the formalisation) depend on the structure-preservation properties of $\mathtt{ren}$, the strength laws $\mathtt{str\text{-}unit}$ and $\mathtt{str\text{-}nat}_2$, and the corollary $\mathtt{str\text{-}dist}$ applied to the coalgebraic map $\mathtt{j} : \mathcal{I} \rightharpoonup ⦅ \mathcal{I} , \mathcal{I} ⦆$.

We therefore have a pointed □-coalgebra instance $\mathbb{T}*□ : \mathbf{Coalg}_* \mathbb{T}$ that is then used as a traversal parameter to induce the substitution map $\mathtt{sub} : \mathbb{T} \rightharpoonup ⦅ \mathbb{T} , \mathbb{T} ⦆$. Initiality once again helps us prove the substitution monoid laws abstractly; in fact, the reasoning steps very closely resemble those needed in the proof of the renaming structure, and they can all be presented in the form of clear, categorical proofs. Once the instances $\mathbb{T}^{\mathrm{M}} : \mathbf{Mon}\ \mathbb{T}$ and $\mathbb{T}^{□\mathrm{M}} : \mathbf{CoalgMon}\ \mathbb{T}$ are derived, we further establish that $\mathbb{T}$ is a Σ-monoid, the proof of which relies on the fact that coalgebraic monoids identify the given coalgebra instance $\mathbb{T}*□$ with the one induced from substitution.

The effort put into identifying the notions of coalgebraic map and strength (Section 2.3.2) pays off repeatedly: the comultiplication law, coalgebraic axioms, and substitution associativity (the *fusion lemmas* of Allais et al. [2017, Section 9.2]) are all established using $\mathtt{str\text{-}dist}$, instantiated at different coalgebraic maps (the application map $\mathtt{j}$, renaming $\mathtt{ren}$ and substitution $\mathtt{sub}$) to derive the generalised forms of the four lifting laws listed by Benton et al. [2012]. The renaming and substitution operations and their correctness laws are derived in an elegant, mathematically-motivated manner, with no auxiliary definitions and ad-hoc lemmas, or additional reasoning machinery needed (such as the bisimulation/fusion framework developed by AACMM [2021]).

### 3.5 Free Σ-Monoid Structure

In the previous section we generically derived a Σ-monoid structure for $\mathbb{T}\,𝔛$, for any family of metavariables $𝔛$. This gives us a lawful substitution operation on $\mathbb{T}\,𝔛$ that can be used in further developments, such as an operational semantics or equational theory. We can go further, however, and characterise $\mathbb{T}\,𝔛$ as the *free Σ-monoid on* $𝔛$: given any Σ-monoid $\mathcal{M}$ and metavariable interpretation $\omega : 𝔛 \rightharpoonup \mathcal{M}$, there is a unique Σ-monoid homomorphism $\omega^{\#} : \mathbb{T}\,𝔛 \rightharpoonup \mathcal{M}$ that satisfies $\omega^{\#}(\mathtt{mvar}\ 𝔪\ \mathtt{var}) = \omega(𝔪)$ for every metavariable $𝔪 : 𝔛\ \tau\ \Pi$. As expected, $\omega^{\#}$ is constructed by initiality, using the fact that the $\overline{\Sigma}$-monoid $\mathcal{M}$ is a meta-algebra with metavariable operator $\mathfrak{X} \xrightarrow{\omega} \mathcal{M} \xrightarrow{\mu} (\!(\mathcal{M}, \mathcal{M})\!)$. Initiality is also used to establish that the unique map $\omega^{\#}$ preserves substitution, i.e. that it is a $\overline{\Sigma}$-monoid homomorphism.

Freeness induces a free-forgetful adjunction between the categories of sorted families and of $\overline{\Sigma}$-monoids, and makes $\mathbb{T}$ into the *free $\overline{\Sigma}$-monoid monad* on sorted families, whose Kleisli extension $(\mathfrak{X} \multimap \mathbb{T}\,\mathfrak{Y}) \to (\mathbb{T}\,\mathfrak{X} \multimap \mathbb{T}\,\mathfrak{Y})$ acts as a form, albeit limited, of *metasubstitution* [Hamana 2004]: occurrences of metavariables from a family $\mathfrak{X}$ in a term of $\mathbb{T}\,\mathfrak{X}$ get replaced with terms of $\mathbb{T}\,\mathfrak{Y}$ according to a mapping $\mathfrak{X} \multimap \mathbb{T}\,\mathfrak{Y}$. This important notion and its generalisation are discussed next.

### 3.6 Metasubstitution

A main difference between metasubstitution and object-level substitution is that the former is *capture-permitting*: a metavariable $\mathfrak{m}$ in a term $\lambda x : \mathbb{N}. \mathfrak{m}\langle x\rangle$ can be replaced with terms that contain free occurrences of $x$; so that valid instances could be $\lambda x.\, x$, $\lambda x.\, x \cdot (x+1)$, or $\lambda x.\, \mathfrak{p}\langle x, x\rangle$ for a metavariable $\mathfrak{p} : \mathfrak{B}\,\mathbb{N}\,[\mathbb{N}, \mathbb{N}]$. The aforementioned metasubstitution map $(\mathfrak{X} \multimap \mathbb{T}\,\mathfrak{Y}) \to (\mathbb{T}\,\mathfrak{X} \multimap \mathbb{T}\,\mathfrak{Y})$ is limited in that the only free variables that the mapping $\zeta : \mathfrak{X} \multimap \mathbb{T}\,\mathfrak{Y}$ may refer to are the parameters of the metavariable: for example, in the open term $x : \mathbb{N}, y : \mathbb{N} \vdash \mathfrak{m}\langle y\rangle$ one cannot instantiate $\mathfrak{m}$ with $x + y$, since $x$ is not a parameter of $\mathfrak{m}$ and is therefore not in scope in the output of $\zeta$. As we wish metasubstitution to model "textual replacement", this is an unnatural restriction.

Our goal then is to capture the following intuition [Fiore 2008]: the term that replaces a metavariable $\mathfrak{m}$ can feature both the parameters of $\mathfrak{m}$, and all the object-level variables in scope at the occurrence of $\mathfrak{m}$ in the term. In elementary terms, the type of the operation may be given as

$$\textsf{msub} : \forall\{\alpha\,\Gamma\} \to (t : \mathbb{T}\,\mathfrak{X}\,\alpha\,\Gamma) \to (\zeta : \forall\{\tau\,\Pi\} \to \mathfrak{X}\,\tau\,\Pi \to \mathbb{T}\,\mathfrak{Y}\,\tau\,(\Pi + \Gamma)) \to \mathbb{T}\,\mathfrak{Y}\,\alpha\,\Gamma$$

As usual, we aim to represent this operation as a morphism of families, and ideally derive it by initiality. Key to this is recognising the type $\forall\{\Pi\} \to \mathfrak{X}\,\tau\,\Pi \to \mathbb{T}\,\mathfrak{Y}\,\tau\,(\Pi + \Gamma)$ as the linear exponential of families $(\mathfrak{X} \multimap \mathbb{T}\,\mathfrak{Y})\,\tau\,\Gamma$ (Section 2.1.2). The derived definition $[\,\mathcal{X} \multimap \mathcal{Y}\,]\,\Gamma \triangleq \forall\{\tau\} \to (\mathcal{X} \multimap \mathcal{Y})\,\tau\,\Gamma$ combines two sorted families into an unsorted one. We also modify the family exponential to take an unsorted family as first argument; that is, overloading notation: $(X \Rightarrow \mathcal{Y})\,\alpha \triangleq X \Rightarrow (\mathcal{Y}\,\alpha)$. The type of metasubstitution may be now succinctly expressed as

$$\textsf{msub} : \mathbb{T}\,\mathfrak{X} \multimap ([\mathfrak{X} \multimap \mathbb{T}\,\mathfrak{Y}] \Rightarrow \mathbb{T}\,\mathfrak{Y})$$

We can then use the following general result to induce $\textsf{msub}$ by initiality: given a $\overline{\Sigma}$-monoid $\mathcal{M}$, the family $[\,\mathfrak{X} \multimap \mathcal{M}\,] \Rightarrow \mathcal{M}$ acquires a $(\overline{\Sigma}, \mathfrak{X})$-meta-algebra structure. However, this only applies if we assume an additional property of the signature endofunctor $\overline{\Sigma}$: it comes with an *exponential strength* $\textsf{estr} : \overline{\Sigma}(X \Rightarrow \mathcal{Y}) \multimap (X \Rightarrow \overline{\Sigma}\,\mathcal{Y})$ (equivalent to the usual Cartesian strength) for every unsorted $\square$-coalgebra $X$ (i.e. unsorted family $X$ with an operation $X\,\Gamma \to (\Gamma \rightsquigarrow \Delta) \to X\,\Delta$). With this in place, the meta-algebra structure on $[\,\mathfrak{X} \multimap \mathcal{M}\,] \Rightarrow \mathcal{M}$ is given as follows:

$$\textsf{MSub}^{\Sigma} : (\mathfrak{X} : \textsf{Family}_s) \to \Sigma\textsf{Mon}\,\mathcal{M} \to \textsf{MetaAlg}\,\mathfrak{X}\,([\,\mathfrak{X} \multimap \mathcal{M}\,] \Rightarrow \mathcal{M})$$

$$\textsf{MSub}^{\Sigma}\,\mathfrak{X}\,\mathcal{M}^{\Sigma M} = \textsf{record}\,\{\,\textsf{alg} \quad = \lambda\,t \quad \zeta \to \mathcal{M}.\textsf{alg}\,(\textsf{estr}\,[\,\mathfrak{X} \multimap \mathcal{M}.^{\square}\,]^{\square}\,M\,t\,\zeta)$$
$$;\,\textsf{var} \quad = \lambda\,v \quad \zeta \to \mathcal{M}.\eta\,v$$
$$;\,\textsf{mvar} = \lambda\,\mathfrak{m}\,\varepsilon\,\zeta \to \mathcal{M}.\mu\,(\zeta\,\mathfrak{m})\,(\textsf{copair}\,\mathcal{M}\,(\lambda\,v \to \varepsilon\,v\,\zeta)\,\mathcal{M}.\eta)\,\}$$

- For $\textsf{alg}$, we use $\textsf{estr}$ to swap the $\overline{\Sigma}$-algebra structure and dependence on the metasubstitution map. We make use of the fact that $[\,\mathfrak{X} \multimap \mathcal{P}\,]$ is an unsorted coalgebra if $\mathcal{P}$ is a sorted coalgebra.
- For $\textsf{var}$, we ignore $\zeta$ entirely and use the point of $\mathcal{M}$.
- For $\textsf{mvar}$, we start by looking up the metavariable $\mathfrak{m} : \mathfrak{X}\,\tau\,\Pi$ in the linear metasubstitution map $\zeta : [\,\mathfrak{X} \multimap \mathcal{M}\,]\,\Gamma$, obtaining the term $\zeta\,\mathfrak{m} : \mathcal{M}\,\tau\,(\Pi + \Gamma)$ in an extended context. We need to tweak this further to get the required output in $\mathcal{M}\,\tau\,\Gamma$, which we do by applying a substitution $(\Pi + \Gamma) -[\, \mathcal{M} \,] \to \zeta$ $\mathfrak{m}$. It is constructed as the copairing of the unit $\mathcal{M}.\eta : \Gamma -[\, \mathcal{M} \,] \to \Gamma$ and the substitution map $\Pi -[\, \mathcal{M} \,] \to \Gamma$ which looks up a variable in context $\Pi$ in $\mathfrak{m}$'s metavariable environment $\varepsilon : \Pi -[\, ([\, \mathfrak{X} \multimap \mathcal{P} \,] \Rightarrow \mathcal{M}) \,] \to \Gamma$ and applies the resulting parametrised term to $\zeta$.

This latter specification is quite a mouthful: metasubstitution (which is derived by initiality into the meta-algebra $\mathsf{MSub}^{\Sigma} \mathfrak{X} \, \mathbb{T}^{\Sigma \mathrm{M}}$) is less a form of "textual replacement" and more an intricate surgical procedure with several interconnected parts. The metavariable-preservation property of $\mathsf{msub}$ illuminates the role of recursively applying metasubstitution to the elements of the metavariable environment $\varepsilon$, then substituting the terms into the parameters of $\zeta$ $\mathfrak{m}$:

$$\mathsf{msub} \; (\mathsf{mvar} \; \mathfrak{m} \; \varepsilon) \; \zeta \equiv \mathsf{sub} \; (\zeta \; \mathfrak{m}) \; (\mathsf{copair} \; (\mathbb{T}\mathfrak{Y})) \; (\lambda \, v \to \mathsf{msub} \; (\varepsilon \, v) \; \zeta) \; \mathsf{var})$$

As an illustration of metasubstitution, consider the open term $x : \mathbb{N} \vdash \lambda y : \mathbb{N}. \; \mathfrak{a}\langle x + 1, \mathfrak{b}\langle y \rangle \rangle : \mathbb{N}$ and the metasubstitution mapping $\zeta = (\mathfrak{a}\langle m, n \rangle \mapsto \mathfrak{c}\langle m \rangle \times n; \mathfrak{b}\langle m \rangle \mapsto \mathfrak{c}\langle m + x \rangle)$ in the global context $x : \mathbb{N}$ (note that the latter term includes both the parameter $m$ and the free variable $x$). The evaluation of the metasubstitution proceeds as follows (where $\mathsf{sub} \; t \; [\cdots]$ and $\mathsf{msub} \; t \; \langle\!\langle \cdots \rangle\!\rangle$ denote substitution and metasubstitution into $t$, respectively):

$$
\begin{aligned}
& \mathsf{msub} \; (\lambda y : \mathbb{N}. \; \mathfrak{a}\langle x + 1, \mathfrak{b}\langle y \rangle \rangle) \; \langle\!\langle \zeta \rangle\!\rangle \\
\equiv \; & \lambda y : \mathbb{N}. \; \mathsf{msub} \; (\mathfrak{a}\langle x + 1, \mathfrak{b}\langle y \rangle \rangle) \; \langle\!\langle \mathsf{wk} \; \zeta \rangle\!\rangle && \text{①} \\
\equiv \; & \lambda y : \mathbb{N}. \; \mathsf{sub} \; (\mathfrak{c}\langle m \rangle \times n) \; [m \mapsto x + 1, n \mapsto \mathsf{msub} \; (\mathfrak{b}\langle y \rangle) \; \langle\!\langle \mathsf{wk} \; \zeta \rangle\!\rangle] && \text{②} \\
\equiv \; & \lambda y : \mathbb{N}. \; \mathsf{sub} \; (\mathfrak{c}\langle m \rangle \times n) \; [m \mapsto x + 1, n \mapsto \mathsf{sub} \; (\mathfrak{c}\langle m + x \rangle) \; [m \mapsto y]] && \text{③} \\
\equiv \; & \lambda y : \mathbb{N}. \; \mathsf{sub} \; (\mathfrak{c}\langle m \rangle \times n) \; [m \mapsto x + 1, n \mapsto \mathfrak{c}\langle y + x \rangle] && \text{④} \\
\equiv \; & \lambda y : \mathbb{N}. \; \mathfrak{c}\langle x + 1 \rangle \times \mathfrak{c}\langle y + x \rangle && \text{⑤}
\end{aligned}
$$

In step ① we push the metasubstitution under the binder. A crucial component of this step is the weakening of the terms in the metasubstitution mapping $\zeta$ represented here by $\mathsf{wk} \; \zeta$. Indeed, since the local context changes from $x : \mathbb{N}$ to $y : \mathbb{N}, x : \mathbb{N}$, the de Bruijn index of the variable $x$ has to be shifted without altering the parameters $m, n$ (for example, $m : \mathbb{N}, x : \mathbb{N} \vdash \mathfrak{c}\langle m + x \rangle$ is renamed to $m : \mathbb{N}, y : \mathbb{N}, x : \mathbb{N} \vdash \mathfrak{c}\langle m + x \rangle$ by shifting the index of $x$). Although this modification makes no difference with the named-variable representation displayed here, in practice it becomes an explicit application of $\mathsf{ren}$ – implemented as part of the exponential strength for $\delta$. It is worth noting the delicate interplay between $\mathsf{ren}$, $\mathsf{sub}$, and $\mathsf{msub}$. In step ② we apply the metasubstitution to $\mathfrak{a}$ by looking up the term $\mathfrak{c}\langle m \rangle \times n$, substituting the contents of $\mathfrak{a}$'s metavariable environment for $m$ and $n$, and recursively metasubstituting $\mathsf{wk} \; \zeta$ into $\mathfrak{b}\langle y \rangle$ (and into $x + 1$, where it is a no-op). The mappings of variables $x, y$ to themselves are omitted. Steps ③ and ④ evaluate the recursive calls, applying an object-level substitution to $\mathfrak{c}\langle m + x \rangle$ that replaces the parameter $m$ with the variable $y$ (and the global variable $x$ with itself). The final substitution is evaluated at step ⑤.

## 3.7 Equational Systems

An immediate application of metasubstitution is building generic *equational systems* for second-order languages [Fiore and Hur 2010; Fiore and Mahmoud 2010]. By specifying the axioms of an equational theory with the use of metavariables, one can use metasubstitution to extract axiom instances between terms and apply rewrite rules within compound expressions. For example, $\beta$-equivalence arises from instances of the axiom

$$\mathfrak{b} : [\alpha]\beta, \; \mathfrak{a} : []\alpha \; \triangleright \; \emptyset \; \vdash \; (\lambda x : \alpha. \; \mathfrak{b}\langle x \rangle) \; \mathfrak{a}\langle \rangle \; \approx \; \mathfrak{b}\langle \mathfrak{a}\langle \rangle \rangle : \beta$$

with terms (of the appropriate types and contexts) metasubstituted for the metavariables $\mathfrak{a}$ and $\mathfrak{b}$. Generic equality can be then directly encoded in Agda as the smallest equivalence relation closed under a given relation $\mathsf{Axiom} : \forall (\mathfrak{X} \; \Gamma \; \{\alpha\}) \to \mathbb{T} \; \mathfrak{X} \; \alpha \; \Gamma \to \mathbb{T} \; \mathfrak{X} \; \alpha \; \Gamma \to \mathsf{Set}$ and metasubstitution:

```
data _▷_⊢_≈_ : (𝔛 : Familyₛ)(Γ : Ctx){α : T} → 𝕋 𝔛 α Γ → 𝕋 𝔛 α Γ → Set₁ where
  eq  : t ≡ s → 𝔛 ▷ Γ ⊢ t ≈ s
  sy  : 𝔛 ▷ Γ ⊢ t ≈ s → 𝔛 ▷ Γ ⊢ s ≈ t
  tr  : 𝔛 ▷ Γ ⊢ t ≈ s → 𝔛 ▷ Γ ⊢ s ≈ u → 𝔛 ▷ Γ ⊢ t ≈ u
  ax  : Axiom 𝔛 Γ t s → 𝔛 ▷ Γ ⊢ t ≈ s
  ms  : 𝔛 ▷ Γ ⊢ t ≈ s → (ζ ξ : [ 𝔛 ⊸ 𝕋 𝔜 ] Γ) →
        (∀{τ Π}(m : 𝔛 τ Π) → 𝔜 ▷ Π ⊹ Γ ⊢ ζ m ≈ ξ m) → 𝔜 ▷ Γ ⊢ msub t ζ ≈ msub s ξ
```

The ms constructor expresses that whenever two terms $t$ and $s$ are equivalent, and two instantiations for their metavariable context $\zeta$ and $\xi$ are equivalent (for every metavariable), then performing the metasubstitution also gives equivalent terms. Using the equivalence constructors one can derive useful proof combinators and a library for equational reasoning; for example, ax≈ equates two terms via the instantiation of an axiom:

$$ax\approx : Axiom\ 𝔛\ Γ\ t\ s → (ζ : [\ 𝔛 ⊸ 𝕋\ 𝔜\ ]\ Γ) → 𝔜 ▷ Γ ⊢ msub\ t\ ζ ≈ msub\ s\ ζ$$
$$ax\approx a\ ζ = ms\ (ax\ a)\ ζ\ ζ\ (λ\ \_{→}\ eq\ refl)$$

The biggest gains, however, come from not having to tediously encode the congruence rules for every subterm of every term of the syntax. To rewrite a deeply nested subexpression, we simply mark its location in the term with a "typed hole" implemented as a distinguished metavariable ◯, and use ms to instantiate it with the two sides of an equality rule: for example, $f ≈ g$ and $(◯\ a) ≈ (◯\ a)$ imply that $f\ a = $ msub $(◯\ a)\ (λ\{◯ → f\}) \overset{ms}{≈} $ msub $(◯\ a)\ (λ\{◯ → g\}) = g\ a$. Further examples of this and other combinators are given in Section 5. An important future development is generically proving the soundness and completeness of second-order equational logic.

We thus conclude our abstract development of initial algebra semantics and move on to the discussion of second-order signatures.

## 4  GENERIC SIGNATURES

The abstract development discussed so far was entirely generic over the second-order signature and term syntax. In this section we discuss how endofunctors $\overline{\Sigma}$ are constructed from descriptions of syntax signatures, the benefits and drawbacks of various term representations, and how we leverage code generation to turn our library into a practical framework for language formalisation. Thanks to our modular implementation, we have several choices in each of the following matters:

- how to encode the signature of a second-order syntax (e.g. binding algebras [Fiore et al. 1999], indexed containers [Altenkirch et al. 2015], AACMM [2021]-style Descriptions);
- how to convert the signature into a **Famₛ** endofunctor $\overline{\Sigma}$ (e.g. polynomial functors [Fiore 2012; Arkor and Fiore 2020], higher- or first-order argument collections, Desc interpretations);
- how to define the data type for the initial $(\overline{\Sigma}, 𝔛)$-meta-algebra (implicit or explicit encodings).

Each of these have their benefits and drawbacks, and we identify three choices that work particularly well together and combine convenience, flexibility, and appropriate computational behaviour.

### 4.1  Binding Signatures

*Binding signatures* were introduced by Aczel [1978] as a generalisation of standard algebraic signatures to languages with variable binding. Our formalisation will use the typed variant of the notion as given in [Fiore and Hur 2010]:

*Definition 4.1.* A *second-order signature* $\Sigma = (T, O, |{-}|)$ is specified by a set of sorts $T$, a set of operators $O$, and an arity function $|{-}| : O → $ List $((\text{List } T) \times T) \times T$.

For an operator o, the tuple $|o| = ([(\vec{\alpha_1}, \beta_1), (\vec{\alpha_2}, \beta_2), \ldots, (\vec{\alpha_n}, \beta_n)], \tau)$ consists of the output sort $\tau \in T$, and a list of $n$ arguments of type $\beta_i$, with each argument binding variables as listed in $\vec{\alpha_i}$. We write this more concisely as o: $[\vec{\alpha_1}]\beta_1, \ldots, [\vec{\alpha_n}]\beta_n \to \tau$, omitting empty bound variable lists.

*Example 4.2.* The second-order signature of the simply-typed $\lambda$-calculus over a base type has the set of types $T$ generated inductively from a base type $N$ and a binary function type $\succ$, and the two type-indexed families of operators

$$\mathsf{app}_{\alpha,\beta} \; : \; (\alpha \succ \beta), \; \alpha \; \to \; \beta \qquad \mathsf{lam}_{\alpha,\beta} \; : \; [\alpha]\beta \; \to \; (\alpha \succ \beta)$$

This definition of signatures can be adapted to Agda almost verbatim, using Ctx in place of List $T$:

```
record Signature (O : Set) : Set₁ where        Arity : O → List (Ctx × T)
  constructor sig                              Arity o = proj₁ | o |
  field |_| : O → List (Ctx × T) × T
                                               Sort : O → T
                                               Sort o = proj₂ | o |
```

The set $T$ of types (or sorts) is normally given as an inductive data type, and $O$ as an enumeration of operators. For example, the STLC has the following sorts and operator symbols:

```
data ΛT : Set where                    data Λₒ : Set where
  N   : ΛT                               appₒ lamₒ : {α β : ΛT} → Λₒ
  _≻_ : ΛT → ΛT → ΛT
```

The Signature instances are direct translations of Example 4.2 above. One can use some simple shorthands for specifying arguments and bound variables to make the declaration concise.

```
Λ:Sig : Signature Λₒ
Λ:Sig = sig λ where (appₒ {α}{β}) → (⊢₀ α ≻ β) , (⊢₀ α)   ↦₂   β
                    (lamₒ {α}{β}) →      (α ⊢₁ β)         ↦₁   α ≻ β
```

## 4.2 Signature Endofunctor

The signature contains all the information needed to determine the syntactic structure of a language. To make use of the abstract development in Section 3, we need to convert a Signature into a sorted-family endofunctor $\bar{\Sigma}$, which captures the way in which constructors of the syntax are associated with arguments. For example, given the signature of the STLC, an element of $\bar{\Sigma} \, \mathcal{X} \, \beta \, \Gamma$ may be the operator app associated with two $\mathcal{X}$-terms $f : \mathcal{X} \, (\alpha \succ \beta) \, \Gamma$ and $a : \mathcal{X} \, \alpha \, \Gamma$, while an element of $\bar{\Sigma} \, \mathcal{X} \, (\alpha \succ \beta) \, \Gamma$ may be the operator lam with a term $b : \mathcal{X} \, \beta \, (\alpha \cdot \Gamma)$.

For technical reasons, that we will expand upon later, we choose to represent the "collection" of arguments of an operator as a tuple of terms. An alternative would be a higher-order encoding as a mapping from an argument index to a term (similar to the implementation of substitutions as context maps); however, even though constructing the Strength for such a representation would be easier, it complicates the initiality proof which we wish to keep as simple as possible.

```
Arg : List (Ctx × T) → Familyₛ → Family
Arg []             𝒳 Γ = ⊤
Arg ((Θ , τ) :: as) 𝒳 Γ = δ Θ 𝒳 τ Γ × Arg as 𝒳 Γ
```

Note the use of the context extension endofunctor $\delta$: it is used to add the new variables bound by an argument to the the global context, making all variables available in the body of the binder.

We are ready to give the definition of the signature endofunctor for a signature $(T, O, \text{Arity}, \text{Sort})$:

$$\Sigma : \text{Family}_{\text{S}} \to \text{Family}_{\text{S}}$$
$$\Sigma \, \mathcal{X} \, \alpha \, \Gamma = \Sigma[\, o \in O \,] \, (\alpha \equiv \text{Sort} \, o \times \text{Arg} \, (\text{Arity} \, o) \, \mathcal{X} \, \Gamma)$$

An element of the set $\Sigma \, \mathcal{X} \, \alpha \, \Gamma$ is a dependent tuple consisting of an operator symbol $o \in O$, a proof that the output sort of the operator is $\alpha$, and a tuple of $\mathcal{X}$-terms for each operand of the operator of the type and extension context given by the operator arity. For example, an element of $\Sigma \, \mathcal{X} \, \beta \, \Gamma$ is $(\text{app}, \text{refl}, (f, t, \text{tt}))$, for terms $f : \mathcal{X} \, (\alpha \rightarrowtail \beta) \, \Gamma$ and $t : \mathcal{X} \, \alpha \, \Gamma$. One can suppress the $\text{tt}$ for operators of positive arity by adding a case for a singleton argument list in the definition of $\text{Arg}$, and use a pattern synonym [Pickering et al. 2016] to hide the $\text{refl}$ element, writing $\text{app} \vdots (f, t)$ for the above.

The only other construction we need is the $\text{Strength}$ instance for $\Sigma$. The family $\mathcal{X}$ in a $(\Sigma \, \mathcal{X})$-term is only used in the argument list, so the $\Sigma$-strength can be easily derived from the $\text{Arg}$-strength $\text{Arg} \, as \, \langle\!\langle \, \mathcal{P} \, , \mathcal{X} \, \rangle\!\rangle \rightarrow \langle\!\langle \, \mathcal{P} \, , \text{Arg} \, as \, \mathcal{X} \, \rangle\!\rangle$. Applying strength to the tuple of arguments simply applies it to every component of type $\delta \, \Theta \, \langle\!\langle \, \mathcal{P} \, , \mathcal{X} \, \rangle\!\rangle \, \tau \, \Gamma$ – and this is nothing but the strength instance for $\delta$ which we constructed back in Section 2.3.2.

$\text{str}^A : (\mathcal{P}^{\square}_* : \text{Coalg}_*\, \mathcal{P})(\mathcal{X} : \text{Family}_{\text{S}})(as : \text{List} \, (\text{Ctx} \times T)) \to \text{Arg} \, as \, \langle\!\langle \, \mathcal{P} \, , \mathcal{X} \, \rangle\!\rangle \rightarrow \langle\!\langle \, \mathcal{P} \, , \text{Arg} \, as \, \mathcal{X} \, \rangle\!\rangle$
$\text{str}^A \, \mathcal{P}^{\square}_* \, \mathcal{X} \, [] \qquad\qquad\quad \text{tt} \qquad\quad \sigma = \text{tt}$
$\text{str}^A \, \mathcal{P}^{\square}_* \, \mathcal{X} \, ((\Theta \, , \tau) :: as) \, (h \, , hs) \, \sigma = (\delta{:}\text{Str.str} \, \Theta \, \mathcal{P}^{\square}_* \, \mathcal{X} \, h \, \sigma) \, , (\text{str}^A \, \mathcal{P}^{\square}_* \, \mathcal{X} \, as \, hs \, \sigma)$

The strength laws are similarly established by pointwise application of the appropriate $\delta{:}\text{Str}$ fields to the elements of the argument tuple. Extending $\text{Arg}$-strength to $\Sigma$ is easy, since the operation does not modify the operator or sort equality proof. In addition to $\Sigma{:}\text{Str}$ below, we also have an instance of exponential strength for $\Sigma$ derived via weakening.

$\Sigma{:}\text{Str} : \text{Strength} \, \Sigma\text{F}$
$\Sigma{:}\text{Str} = \text{record} \, \{ \, \text{str} \qquad\quad = \lambda \, \mathcal{P}^{\square}_* \, \mathcal{X} \, (o \, , e \, , a) \, \sigma \to o \, , e \, , (\text{str}^A \, \mathcal{P}^{\square}_* \, \mathcal{X} \, (\text{Arity} \, o) \, a \, \sigma)$
$\qquad\qquad\qquad\quad ; \, \text{str-nat}_1 = \lambda \, f^{\square}_* \Rightarrow \quad (o \, , e \, , a) \, \sigma \to \text{cong} \, (o \, , e \, , \_) \, (\text{str}^A\text{-nat}_1 \, f^{\square}_* \Rightarrow (\text{Arity} \, o) \, a \, \sigma) \, ; ... \}$

## 4.3  Term Syntax

The final piece of the puzzle is constructing the initial $(\Sigma, \mathfrak{X})$-meta-algebra $\mathbb{T}$ from a second-order signature endofunctor. Such initial algebras correspond to inductive data types whose constructors combine $\mathbb{T}$-terms into other $\mathbb{T}$-terms, allowing for arbitrarily nested syntactic structure.

We have several choices in the approach we take. One is to treat the tuples $(\text{op} \vdots (a_1, \ldots, a_n))$ as the terms of the syntax, directly encoding the $\Sigma$-algebra structure as a unified term constructor $\text{con} : \Sigma \, \mathbb{T} \rightarrow \mathbb{T}$; the other is the more common implementation with one constructor for each operator, applicable for signatures with a finite set of operators.

*Implicit Encoding.* Alongside the $\Sigma$-algebraic structure, terms of a second-order syntax also have to include constructors for variables and metavariables. This suggests the following generic data type of terms for an arbitrary signature:

data Tm : $\text{Family}_{\text{S}}$ where                               data Sub $(\mathcal{X} : \text{Family}_{\text{S}}) : \text{Ctx} \to \text{Ctx} \to \text{Set}$ where
    con   : $\Sigma$ Tm $\tau$ $\Gamma$                    $\to$ Tm $\tau$ $\Gamma$               •    : Sub $\mathcal{X}$ $\emptyset$ $\Gamma$
    var   : $\mathcal{I}$ $\tau$ $\Gamma$                        $\to$ Tm $\tau$ $\Gamma$           _◄_ : $\mathcal{X}$ $\alpha$ $\Gamma$ $\to$ Sub $\mathcal{X}$ $\Pi$ $\Gamma$ $\to$ Sub $\mathcal{X}$ $(\alpha \cdot \Pi)$ $\Gamma$
    mvar : $\mathfrak{X}$ $\tau$ $\Pi$ $\to$ Sub Tm $\Pi$ $\Gamma$ $\to$ Tm $\tau$ $\Gamma$

Note the use of Sub (defined on the right above) in place of a context map to represent the metavariable environment. It is a first-order, inductive encoding of a simultaneous substitution of terms in context $\Gamma$ for every variable in context $\Pi$, which, while isomorphic to context maps (with conversion functions lookup and tabulate), is a more appropriate choice for syntax. Recalling that metasubstitution recurses into the metavariable environment, in a higher-order representation the recursive call would get suspended in the body of a $\lambda$-abstraction and lead to terms with unevaluated expressions. With context maps, the application sem (mvar $\mathfrak{m}$ ($\lambda\{$ new $\rightarrow$ con (op $\vdots$ $t$) $\}$)) would only normalise to mvar $\mathfrak{m}$ ($\lambda\{$ new $\rightarrow$ sem (con (op $\vdots$ $t$)) $\}$), and sem would not get pushed further under the constructor con unless the environment function is applied to new. In contrast, using the first-order encoding Sub, each element of the sequence gets fully normalised, so, as desired, sem (mvar $\mathfrak{m}$ (con (op $\vdots$ $t$) ◂ •)) reduces to mvar $\mathfrak{m}$ (con (op $\vdots$ sem $t$) ◂ •). The same reasoning was behind our choice to represent operator arguments as tuples, rather than higher-order assignments – though the strength instance would have been simpler, specifying arguments via case-analysis is cumbersome, and syntactic operations only recurse one layer deep into subterms.

The proof of initiality of Tm asks us to define a recursive function sem : Tm $\rightarrow$ $\mathscr{A}$ for any meta-algebra $\mathscr{A}$, translating constructors of Tm to the algebra and (meta)variable structure of $\mathscr{A}$. There is an issue with the obvious definition: recursively interpreting the subterms of a constructor involves mapping sem over a tuple of terms, which Agda does not recognise as a terminating function call. Allais et al. [2021, Section 4] also encountered this problem and proposed the use of Agda's *sized types* [Abel 2010] (which, unfortunately, are logically unsound in the current Agda version 2.6.2) to mark a term as strictly "larger" than its subterms. The workaround suits their needs of defining sem – at the expense of having to carry around the size index everywhere – but it does not extend to proving the uniqueness of sem which our framework closely relies on (see Pitts [2019] for the technical details of the same issue in the context of general $F$-algebras).

The solution to this issue is surprisingly simple: instead of recursively interpreting the subterms using a functorial mapping $(\mathscr{X} \rightarrow \mathscr{Y}) \rightarrow (\text{Arg } as \, \mathscr{X} \, \Gamma \rightarrow \text{Arg } as \, \mathscr{Y} \, \Gamma)$ (and similarly for Sub), we inline the transformations as the mutually recursive functions $\mathbb{A}$ and $\mathbb{S}$ that apply sem : Tm $\rightarrow$ $\mathscr{A}$ to subterms directly. The termination checker is satisfied without the need for sized types!

$\mathbb{A}$ : $\forall$ $as$ $\rightarrow$ Arg $as$ Tm $\Gamma$ $\rightarrow$ Arg $as$ $\mathscr{A}$ $\Gamma$            $\mathbb{S}$ : Sub Tm $\Pi$ $\Gamma$ $\rightarrow$ $\Pi$ $-[\,\mathscr{A}\,]\rightarrow$ $\Gamma$
$\mathbb{A}$ [ ]            tt           = tt                    $\mathbb{S}$ ($t$ ◂ $\sigma$) new       = sem $t$
$\mathbb{A}$ ($a$ :: $as$) ($t$ , $ts$) = (sem $t$ , $\mathbb{A}$ $as$ $ts$)        $\mathbb{S}$ ($t$ ◂ $\sigma$) (old $v$) = $\mathbb{S}$ $\sigma$ $v$

sem (con ($o$ , $e$ , $a$)) = alg ($o$ , $e$ , $\mathbb{A}$ (Arity $o$) $a$)
sem (var $v$)        = var $v$
sem (mvar $\mathfrak{m}$ $\varepsilon$)    = mvar $\mathfrak{m}$ ($\mathbb{S}$ $\varepsilon$)

The proof that sem is a unique ($\textstyle\sum$, $\mathfrak{X}$)-meta-algebra homomorphism is also quite straightforward, only requiring a few mutually inductive lemmas about $\mathbb{A}$ and $\mathbb{S}$. We then have the following result:

THEOREM 4.3. *Tm is an initial ($\textstyle\sum$, $\mathfrak{X}$)-meta-algebra.*

We can thus instantiate our previous metatheory with this concrete data type to get access to substitution and its correctness properties, sound compositional interpretations in models, etc.

*Explicit Encoding.* The implicit encoding achieves our conceptual goals: it is a first-order initial meta-algebra for an arbitrary second-order signature. Its main practical disadvantage is that the term syntax is closely tied to the algebraic framework and forces users to adopt a cumbersome encoding of terms, rather than the natural and elegant "one constructor for each typing rule" principle of intrinsic typing. Pattern synonyms may be used to simplify the surface syntax, though we found them quite fragile when working with parametrised modules – not to mention that pattern synonyms are untyped and feel inherently "hacky" for something as important as the terms of a formalised language. Ideally, we would like users to be able to adopt our framework as seamlessly as possible, perhaps even plugging it into an existing formalisation without changing the fundamental data type of syntactic terms.

This is very much possible, thanks to our rigid separation between signatures, endofunctors and initial meta-algebras. Defining the initial meta-algebra instance for an existing data type is not an especially laborious task: one needs a recursive initial interpretation function, a homomorphism proof, and an inductive uniqueness proof. Metavariables still require the mutually recursive transformation $\mathbb{S}$ (defined as before and so omitted here), but now that one is manually recursing into subterms, the transformation $\mathbb{A}$ is not needed.

```
data Λ : Familyₛ where                              sem : Λ ⇾ 𝒜
   var  : ℐ τ Γ → Λ τ Γ                             sem (var v)    = var v
   mvar : 𝔛 τ Π → Sub Λ Π Γ → Λ τ Γ                 sem (mvar 𝔪 ε) = mvar 𝔪 (𝕊 ε)
   app  : Λ (α ⇾ β) Γ → Λ α Γ → Λ β Γ               sem (app g a)  = alg (appₒ ∶ sem g , sem a)
   lam  : Λ β (α · Γ) → Λ (α ⇾ β) Γ                 sem (lam b)    = alg (lamₒ ∶ sem b)
```

The homomorphism instance uses a simple lemma $\mathbb{S}$-tab about the interaction of $\mathbb{S}$ and tabulate (for all context maps $\varepsilon$, $\mathbb{S}$ (tabulate $\varepsilon$) = sem ∘ $\varepsilon$), and the $\overline{\sum}$-algebra homomorphism proof, which is satisfied merely by pattern-matching on the operand and sort equality proof.

$$
\begin{aligned}
&\text{sem}^{\Sigma\Rightarrow} : \text{MetaAlg}\Rightarrow Λ^\Sigma \, 𝒜^\Sigma \, \text{sem}\\
&\text{sem}^{\Sigma\Rightarrow} = \text{record } \{ \langle\text{alg}\rangle = λ \{t = t\} → \langle\text{alg}\rangle \; t \; ; \langle\text{var}\rangle = \text{refl}\\
&\qquad\qquad\qquad ; \langle\text{mvar}\rangle = λ \{𝔪 = 𝔪\}\{\varepsilon\} → \text{cong } (\text{mvar } 𝔪) \, (\mathbb{S}\text{-tab } \varepsilon) \}\\
&\quad \text{where } \langle\text{alg}\rangle : (t : \overline{\sum} \, Λ \, α \, Γ) → \text{sem } (Λ^\Sigma.\text{alg } t) \equiv 𝒜^\Sigma.\text{alg } (\overline{\sum}_1 \, \text{sem } t)\\
&\qquad\qquad \langle\text{alg}\rangle \; (\text{app}_\text{o} \; ∶ \; \_) = \text{refl}\\
&\qquad\qquad \langle\text{alg}\rangle \; (\text{lam}_\text{o} \; ∶ \; \_) = \text{refl}
\end{aligned}
$$

The uniqueness proof – that sem is equal to any meta-algebra homomorphism $g : Λ ⇾ 𝒜$ – involves the mutually inductive lemma $\mathbb{S}$-lu and the inverse property of tabulate and lookup in the metavariable case; everything else follows from the homomorphism properties of $g$.

$$
\begin{aligned}
&\mathbb{S}\text{-lu} : (\varepsilon : \text{Sub } Λ \, Π \, Γ)(v : ℐ \, α \, Π) → \mathbb{S} \, \varepsilon \, v \equiv g \, (\text{lookup } \varepsilon \, v)\\
&\mathbb{S}\text{-lu} \; (t ◂ \varepsilon) \; \text{new}     = \text{sem! } t\\
&\mathbb{S}\text{-lu} \; (t ◂ \varepsilon) \; (\text{old } v) = \mathbb{S}\text{-lu } \varepsilon \, v\\[4pt]
&\text{sem!} : (t : Λ \, α \, Γ) → \text{sem } t \equiv g \, t\\
&\text{sem! } (\text{var } v) = \text{sym } \langle\text{var}\rangle\\
&\text{sem! } (\text{mvar } 𝔪 \, \varepsilon) \; \text{rewrite } \mathbb{S}\text{-lu } \varepsilon = \text{trans } (\text{sym } \langle\text{mvar}\rangle) \, (\text{cong } (g ∘ \text{mvar } 𝔪) \, (\text{tab∘lu≈id } \varepsilon))\\
&\text{sem! } (\text{app } f \, a) \quad \text{rewrite sem! } f \mid \text{sem! } a = \text{sym } \langle\text{alg}\rangle\\
&\text{sem! } (\text{lam } b) \qquad \text{rewrite sem! } b \qquad\quad = \text{sym } \langle\text{alg}\rangle
\end{aligned}
$$

At the expense of minimal extra boilerplate, one is able to prove that the inductively defined family $Λ$ is an initial meta-algebra, and instantiate our metatheory with this result. The Theory module associated with an initial meta-algebra exports every definition and lemma given in the framework for easy access; for example, the coveted single-variable substitution operation is directly accessible as Theory.[_/] : $Λ \, α \, Γ → Λ \, β \, (α · Γ) → Λ \, β \, Γ$, with no extra work required.

## 4.4 Code Generation

The next logical step is eliminating the need to write boilerplate altogether and allow users to go from a direct specification of the second-order signature to the formalised metatheory of an explicitly encoded data type of terms automatically. Since the initiality proof for the explicit encoding is rather formulaic and much of it is signature-independent, one may as well generate the associated Agda code from a syntax description using a simple Python script. This signature-to-Agda compiler takes a simple textual specification of a type and term signature, and produces the Agda code for the Signature declaration and initiality proof outlined in Sections 4.1 and 4.3. For instance, one can give the signature of our running STLC example as follows (recall the introduction), also optionally specifying infix symbols and fixity information:

| type | | term | | | |
|---|---|---|---|---|---|
| N | : 0-ary | app | $: \alpha \succ \beta \quad \alpha \quad \rightarrow \beta$ | \| | _$_ l20 |
| _$\succ$_ | : 2-ary \| r30 | lam | $: \alpha.\beta \qquad\quad \rightarrow \alpha \succ \beta$ | \| | $\lambda$_ r10 |

Saving this in a file `stlc` and running `python soas.py stlc` produces the files `Signature.agda` and `Syntax.agda` with all the required imports, declarations of the signature Λ:Sig and explicit inductive data type of terms Λ, and the initiality proof. One is thereby free to proceed with the interesting parts of language formalisation like operational and denotational semantics right away.

The compiler is a fairly straightforward Python program that parses the specification language and produces formatted Agda files using the built-in string templating system. Its simplicity is one of its advantages: the boilerplate code it generates is systematic and minimal (1-2 lines of code for every type constructor and 6-7 lines for every term constructor), so manual testing on a wide range of examples gives us sufficient confidence in the robustness and correctness of the script. Unlike most other code generation solutions (some listed in Section 6.1) that produce the entire syntax-specific formalisation, including fragile and impenetrable de Bruijn arithmetic proofs, we leverage the fully syntax-generic metatheory implemented in the library and generate just enough code (namely the initiality proof, which is less boilerplate and more an elegant categorical argument) to instantiate it. Consequently, the compiler output is concise, readable, and easy to maintain.

Having examined the generic metatheory, construction of signatures, and the term syntax, we conclude with some demonstrations of the framework used in practice.

## 5 EXAMPLES

Our framework is flexible and unopinionated: it can be plugged into any intrinsically-typed formalisation of a second-order calculus and equip the syntax with the often-needed operations of weakening and substitution, and their corresponding laws. It also helps users in defining evaluation functions, interpreters, and syntactic translations in a concise and provably-correct manner. In this section, we give two extended examples of how the library may be used.

### 5.1 Computational Calculi

The STLC has been our running example throughout the paper, and our framework is a great playground for experimenting with various extensions of it, be it with new types, terms, or equations. Below is a list of constructs that can be easily represented and compiled into Agda. The semantics of such a language would be rather complicated, but its syntax is still just a second-order signature.

| type | | term | | | | | | | |
|---|---|---|---|---|---|---|---|---|---|
| N | : 0-ary | app | $: \alpha \succ \beta \quad \alpha$ | $\rightarrow \beta$ | | pair | $: \alpha \quad \beta$ | | $\rightarrow \alpha \otimes \beta$ |
| _$\succ$_ | : 2-ary | lam | $: \alpha.\beta$ | $\rightarrow \alpha \succ \beta$ | | fst | $: \alpha \otimes \beta$ | | $\rightarrow \alpha$ |
| _$\otimes$_ | : 2-ary | | | | | snd | $: \alpha \otimes \beta$ | | $\rightarrow \beta$ |
| _$\oplus$_ | : 2-ary | let | $: \alpha \quad\; \alpha.\beta$ | $\rightarrow \beta$ | | | | | |
| ¬_ | : 1-ary | fix | $: \alpha \succ \alpha$ | $\rightarrow \alpha$ | | inl | $: \alpha$ | | $\rightarrow \alpha \oplus \beta$ |
| T | : 1-ary | | | | | inr | $: \beta$ | | $\rightarrow \alpha \oplus \beta$ |
| | | throw | $: \alpha \qquad \neg\, \alpha$ | $\rightarrow \beta$ | | case | $: \alpha \oplus \beta \quad \alpha.\gamma \quad \beta.\gamma$ | $\rightarrow \gamma$ | |
| | | callcc | $: (\neg\, \alpha).\alpha$ | $\rightarrow \alpha$ | | ze | $:$ | | $\rightarrow$ N |
| | | | | | | su | $:$ N | | $\rightarrow$ N |
| | | return | $: \alpha$ | $\rightarrow$ T $\alpha$ | | nrec | $:$ N $\quad \alpha \quad (\alpha,$N$).\alpha$ | | $\rightarrow \alpha$ |
| | | bind | $:$ T $\alpha \qquad \alpha.($T $\beta)$ | $\rightarrow$ T $\beta$ | | | | | |

We will use the minimal fragment of STLC (with app and lam) to showcase the construction of models and interpretations. The CCC model of the STLC [Lambek 1980] in the category **Set** interprets types as sets, and terms $\Gamma \vdash t : \alpha$ as functions from the interpretation of $\Gamma$ to the interpretation of $\alpha$. Interpretation of contexts can be higher-order or first-order (as a Cartesian product of type interpretations) – with the higher-order encoding the model definition is remarkably concise.

```
⟦_⟧ : ΛT → Set
⟦ N ⟧        = ℕ
⟦ α ▷ β ⟧ = ⟦ α ⟧ → ⟦ β ⟧

⟦_⟧ᶜ : Ctx → Set
⟦ Γ ⟧ᶜ = ∀{α} → 𝓘 α Γ → ⟦ α ⟧

_⁺_ : ⟦ α ⟧ → ⟦ Γ ⟧ᶜ → ⟦ α · Γ ⟧ᶜ
(a ⁺ γ) new   = a
(a ⁺ γ) (old v) = γ v

Env : Familyₛ
Env α Γ = ⟦ Γ ⟧ᶜ → ⟦ α ⟧
```

```
Envᶻᴹ : ΣMon Env
Envᶻᴹ = record
    { ᴹ = record { η = λ v γ → γ v ; μ = λ t σ δ → t (λ v → σ v δ)
                      ; lunit = refl ; runit = refl ; assoc = refl }
    ; alg     = λ { (appₒ ◦ f , a) γ → f γ (a γ)
                      ; (lamₒ ◦ b)        γ → λ a → b (a ⁺ γ) }
    ; μ⟨alg⟩ = λ { (appₒ ◦ _) → refl
                      ; (lamₒ ◦ b) → ext² λ δ a → cong b (dext
                          λ { new → refl ; (old v) → refl }) } }

module Env = FreeMonoid ∅ Envᶻᴹ (λ ())

eval : Λ₀ → Env
eval = Env.𝗌𝖾𝗆
```

Here we restrict to the sorted family $\Lambda_0 = \Lambda\,\varnothing$ of $\lambda$-terms without metavariables for simplicity. The eval function interprets $\lambda$-terms as Agda programs; for example, eval ($\lambda\,\lambda\,x_1$) ($\lambda$ ()) (where the 1st de Bruijn index var (old new) is denoted $x_1$) normalises to the Agda function $\lambda\,x\,y \to x$. Since it is derived by initiality, the interpretation is compositional and satisfies the semantic substitution lemma *by construction*. This is an enormous time-saver, since proving the soundness of substitution is often one of the most tedious steps required for the development of denotational semantics – the binding terms are the usual suspects, forcing one to reason about semantics of lifting, weakening, renaming, etc. Our framework does all the heavy lifting, allowing users to move on to less bureaucratic proofs. For example, after defining the predicate Val satisfied by value terms of the form $\lambda\,b$ for $b : \Lambda_0\,\beta\,(\alpha \cdot \Gamma)$, it takes minimal effort to equip the language with an intrinsically-typed call-by-value reduction relation and a proof that it preserves the interpretation of terms:

```
data _⤳_ : Λ₀ α Γ → Λ₀ α Γ → Set where
    ζ-$₁ : {f g : Λ₀ (α ▷ β) Γ} {a  : Λ₀ α Γ}                      → f ⤳ g → f $ a   ⤳ g $ a
    ζ-$₂ : {f  : Λ₀ (α ▷ β) Γ} {a b : Λ₀ α Γ}       → Val f → a ⤳ b → f $ a   ⤳ f $ b
    β-𝜆 : {t  : Λ₀ α Γ}       {b  : Λ₀ β (α · Γ)} → Val t            → (𝜆 b) $ t ⤳ [ t /] b

        sound : {t s : Λ₀ α Γ} → t ⤳ s → (γ : ⟦ Γ ⟧ᶜ) → eval t γ ≡ eval s γ
        sound (ζ-$₁ r)      γ rewrite sound r γ = refl
        sound (ζ-$₂ _ r)    γ rewrite sound r γ = refl
        sound (β-𝜆 {t}{b} _) γ rewrite Env.sub-lemma t b
            = cong (eval b) (dext λ { new → refl ; (old v) → refl })
```

Note the use of the freely-obtained single-variable substitution [ t /] b in the $\beta$-$\lambda$ axiom, and the invocation of sub-lemma which translates eval ([ t /] b) γ to Env.[ eval t /] (eval s) γ. The STLC is of course a basic calculus, but implementing its syntax, operational and denotational semantics from scratch still involves a considerable amount of effort, mostly spent defining and reasoning about substitution. Leveraging our framework, this standard but relatively robust formalisation is possible in fewer than 100 lines of Agda code. We can also continue in the standard way with proofs like progress, determinacy, normalisation, etc. – if a syntactic property like associativity of substitution is required, it is likely to be present in the Theory module of the library already.

## 5.2 Partial Differentiation

Another example of a second-order calculus is the axiomatisation of partial differentiation laid out by Plotkin [2020]. The syntax consists of the first-order theory of commutative rings with a second-order partial-differentiation operator $\text{PDiff}(x.e\langle x\rangle, d)$, interpreted as the partial derivative of the expression $e\langle x\rangle$ with respect to $x$, evaluated at $d$ (which has no free occurrences of $x$). This is usually denoted $\frac{\partial e\langle x\rangle}{\partial x}\big|_{x=d}$. To differentiate $e\langle x\rangle$ without evaluation, one renames $x$ to a dummy variable $w$, differentiates $e\langle w\rangle$ with respect to $w$, then evaluates the result at $w = x - $ thus, the notation $\frac{\partial}{\partial x}e\langle x\rangle$ is taken as abbreviating $\frac{\partial e\langle w\rangle}{\partial w}\big|_{w=x}$.

The signature of rings augmented with the partial-differentiation operator can be readily expressed as an unsorted syntax description. We manually implement some derived operators in Agda, such as $\partial_0$ and $\partial_1$ (with symbols d0 and d1), respectively the partial derivatives w.r.t. the first and second variable. Note the use of weakening below, available from the Theory module.

```
term
  zero  :                    * | 𝟘
  one   :                    * | 𝟙
  inv   :  *       →         * | ⊖_  r50
  add   :  *   *   →         * | _⊕_ l20
  mult  :  *   *   →         * | _⊗_ l40
  pdiff : *.*  *   →         * | ∂_|_
```

$\partial_{0\_}$ : PD $\mathfrak{X}$ * $(* \cdot \Gamma) \to$ PD $\mathfrak{X}$ $(* \cdot \Gamma)$
$\partial_0\ e = \partial$ (Theory.wk $\mathfrak{X}$ $e$) | $x_0$
$\partial_{1\_}$ : PD $\mathfrak{X}$ * $(* \cdot * \cdot \Gamma) \to$ PD $\mathfrak{X}$ $(* \cdot * \cdot \Gamma)$
$\partial_1\ e = \partial$ (Theory.wk $\mathfrak{X}$ $e$) | $x_1$

The equational theory may also be included in the syntax description, listed both as explicit equations of the form '(name) metavars ▷ vars ⊢ expr$_1$ = expr$_2$', and algebraic properties of operators. Plotkin's use of function variables matches up with parametrised metavariables, so the axioms of the paper can be directly translated to our formalism. A few examples are given below.

```
theory
  'zero' unit of 'add'
  'mult' distributes over 'add'
  (∂⊕) a : *        ▷ x : *       ⊢ d0 (add (x,a))  = one
  (∂C) f : (*,*).*  ▷ x : *  y : * ⊢ d1 (d0 (f[x,y])) = d0 (d1 (f[x,y]))
```

The generated Agda modules include the intrinsically-typed syntax of semirings with partial differentiation (as an inductive sorted family PD), and the generic equational reasoning framework of Section 3.7 instantiated with the data type $\_\triangleright\_\vdash\_\approx_A\_$ generated from the axiom descriptions. It employs some syntactic sugar for the purposes of readability and ease of use. Instead of defining a named inductive family of named metavariables for every axiom, we build up a context of metavariables in-place with the notation $\lceil \Pi_1 \Vdash \tau_1 \rceil \lceil \Pi_2 \Vdash \tau_2 \rceil \cdots \lceil \Pi_n \Vdash \tau_n \rceil$, and refer to the metavariables using (alphabetic) de Bruijn indices $\mathfrak{a}, \mathfrak{b}, \mathfrak{c}$. The environment for an $n$-ary metavariable $\mathfrak{m}$ is specified as $\mathfrak{m}\langle t_0 \blacktriangleleft \ldots \blacktriangleleft t_{n-1} \rangle$, or just $\mathfrak{m}$ if $n = 0$. Some examples are shown below:

data $\_\triangleright\_\vdash\_\approx_A\_$ : $(\mathfrak{X} : \text{MCtx})(\Gamma : \text{Ctx})\{a : *\mathsf{T}\} \to$ PD $\underline{\mathfrak{X}}\ \alpha\ \Gamma \to$ PD $\underline{\mathfrak{X}}\ \alpha\ \Gamma \to$ Set where
  𝟘U⊕$^L$  : $\lceil * \rceil$                        ▷ $\emptyset$       ⊢ $0 \oplus \mathfrak{a}$                      $\approx_A \mathfrak{a}$
  ⊗D⊕$^L$  : $\lceil * \rceil \lceil * \rceil \lceil * \rceil$ ▷ $\emptyset$  ⊢ $\mathfrak{a} \otimes (\mathfrak{b} \oplus \mathfrak{c})$           $\approx_A (\mathfrak{a} \otimes \mathfrak{b}) \oplus (\mathfrak{a} \otimes \mathfrak{c})$
  ∂⊕      : $\lceil * \rceil$                  ▷ $\lfloor * \rfloor$ ⊢ $\partial_0\ (x_0 \oplus \mathfrak{a})$            $\approx_A \mathbb{1}$
  ∂⊗      : $\lceil * \rceil$                  ▷ $\lfloor * \rfloor$ ⊢ $\partial_0\ (\mathfrak{a} \otimes x_0)$            $\approx_A \mathfrak{a}$
  ∂C      : $\lceil * \cdot * \Vdash * \rceil$     ▷ $\lfloor * \cdot * \rfloor$ ⊢ $\partial_1\ (\partial_0\ \mathfrak{a}\langle x_0 \blacktriangleleft x_1 \rangle) \approx_A \partial_0\ (\partial_1\ \mathfrak{a}\langle x_0 \blacktriangleleft x_1 \rangle)$
  ∂Ch2    : $\lceil * \cdot * \Vdash * \rceil \lceil * \Vdash * \rceil \lceil * \Vdash * \rceil$ ▷ $\lfloor * \rfloor$ ⊢
            $\partial_0\ \mathfrak{a}\langle\ \mathfrak{b}\langle x_0 \rangle \blacktriangleleft \mathfrak{c}\langle x_0 \rangle\ \rangle \approx_A (\partial\ \mathfrak{a}\langle x_0 \blacktriangleleft \mathfrak{c}\langle x_1 \rangle\ \rangle\ |\ \mathfrak{b}\langle x_0 \rangle) \otimes (\partial_0\ \mathfrak{b}\langle x_0 \rangle) \oplus$
            $(\partial\ \mathfrak{a}\langle\ \mathfrak{b}\langle x_1 \rangle \blacktriangleleft x_0 \rangle\ |\ \mathfrak{c}\langle x_0 \rangle) \otimes (\partial_0\ \mathfrak{c}\langle x_0 \rangle)$

The first two constructors correspond to the left unit and distributivity laws, with the right versions automatically derived via the commutativity of $\oplus$ and $\otimes$ (omitted). The four constructors involving differentiation correspond to the axioms of sums and products ($\frac{\partial x \oplus a}{\partial x} = 1$, $\frac{\partial a x}{\partial x} = a$), exchange of derivative operators ($\frac{\partial}{\partial y} \frac{\partial}{\partial x} f(x, y) = \frac{\partial}{\partial x} \frac{\partial}{\partial y} f(x, y)$), and the binary version of the chain rule. Note that these axioms relate open terms with one or two object variables, which must be listed in the context in addition to the metavariables.

We may now feed the axiom relation to the signature-generic equational logic and establish theorems via equational reasoning – be they properties of rings, derivable partial-differentiation laws, or explicit calculations of derivatives. The user can once again make use of some syntactic sugar to streamline the proofs. For a simple example, consider the corollary $\frac{\partial 0}{\partial x} = 0$, derivable from the left annihilation law of $0$ and $\otimes$, and the product differentiation axiom.

$$\partial 0 : [\,] \rhd \lfloor \ast \rfloor \vdash \partial_0\, 0 \approx 0$$
$$\partial 0 = \text{begin}\quad \partial_0\, 0 \qquad\qquad\qquad \approx \langle\ \text{cong}[\ \text{ax } 0\text{X}\otimes^L \text{ with } \langle\!\langle\ \text{x}_0\ \rangle\!\rangle\ ]\text{in } \partial_0\, \bigcirc^{\text{a}}\ \rangle_{\text{s}}$$
$$\partial_0\, (0 \otimes \text{x}_0) \qquad \approx \langle\ \text{ax } \partial\otimes \text{ with } \langle\!\langle\ 0\ \rangle\!\rangle\ \rangle$$
$$0 \qquad\qquad\qquad \blacksquare$$

The $\text{ax } a$ with $\langle\!\langle\ t_1 \lhd \ldots \lhd t_n\ \rangle\!\rangle$ notation associates an axiom $a$ with an instantiation of metavariables in the metavariable context of the axiom, given as a list of terms. Applications of an equation in a subexpression of $t$ is done with the $\text{cong}[\ e\ ]\text{in } t\langle\mathfrak{m}\rangle$ combinator, where $\mathfrak{m}$ is a new metavariable (denoted $\bigcirc^{\text{m}}$ to make its role as a 'hole' clear) added to the context to indicate the location in which the equation $e$ is applied. Thus, in the first step, we apply the left annihilation axiom $0 = 0 \otimes x$ instantiated at $x = \text{x}_0$ to the subexpression of the $\partial_0$ operator. As a more involved example, we derive the unary chain rule from the binary axiom (instantiated with $f(x, y) \triangleq f(x)$ and $h(x) = 0$), the $\partial 0$ corollary above, and the unit and annihilation laws for $0$:

$$\partial\text{Ch}_1 : [\ast \Vdash \ast\ ]\ [\ \ast \Vdash \ast\ ] \vdash \partial_0\, \mathfrak{a}\langle\, \mathfrak{b}\langle\, \text{x}_0\, \rangle\, \rangle \approx (\partial\, \mathfrak{a}\langle\, \text{x}_0\, \rangle\ |\ \mathfrak{b}\langle\, \text{x}_0\, \rangle) \otimes (\partial_0\, \mathfrak{b}\langle\, \text{x}_0\, \rangle)$$
$$\partial\text{Ch}_1 = \text{begin}\quad \partial_0\, \mathfrak{a}\langle\, \mathfrak{b}\langle\, \text{x}_0\, \rangle\, \rangle$$
$$\approx \langle\ \text{ax } \partial\text{Ch}_2 \text{ with } \langle\!\langle\ \mathfrak{a}\langle\, \text{x}_0\, \rangle \lhd \mathfrak{b}\langle\, \text{x}_0\, \rangle \lhd 0\ \rangle\!\rangle\ \rangle$$
$$(\partial\, \mathfrak{a}\langle\, \text{x}_0\, \rangle\ |\ \mathfrak{b}\langle\, \text{x}_0\, \rangle) \otimes (\partial_0\, \mathfrak{b}\langle\, \text{x}_0\, \rangle) \oplus (\partial\, \mathfrak{a}\langle\, \mathfrak{b}\langle\, \text{x}_1\, \rangle\, \rangle\ |\ 0) \otimes \partial_0\, 0$$
$$\approx \langle\ \text{cong}[\ \text{thm } \partial 0\ ]\text{in } [\ldots] \oplus (\partial\, \mathfrak{a}\langle\, \mathfrak{b}\langle\, \text{x}_1\, \rangle\, \rangle\ |\ 0) \otimes \bigcirc^{\text{c}}\ \rangle$$
$$(\partial\, \mathfrak{a}\langle\, \text{x}_0\, \rangle\ |\ \mathfrak{b}\langle\, \text{x}_0\, \rangle) \otimes (\partial_0\, \mathfrak{b}\langle\, \text{x}_0\, \rangle) \oplus (\partial\, \mathfrak{a}\langle\, \mathfrak{b}\langle\, \text{x}_1\, \rangle\, \rangle\ |\ 0) \otimes 0$$
$$\approx \langle\ \text{cong}[\ \text{thm } 0\text{X}\otimes^R \text{ with } \langle\!\langle\ \partial\, \mathfrak{a}\langle\, \mathfrak{b}\langle\, \text{x}_1\, \rangle\, \rangle\ |\ 0\ \rangle\!\rangle\ ]\text{in } [\ldots] \oplus \bigcirc^{\text{c}}\ \rangle$$
$$(\partial\, \mathfrak{a}\langle\, \text{x}_0\, \rangle\ |\ \mathfrak{b}\langle\, \text{x}_0\, \rangle) \otimes (\partial_0\, \mathfrak{b}\langle\, \text{x}_0\, \rangle) \oplus 0$$
$$\approx \langle\ \text{ax } 0\text{U}\oplus^R \text{ with } \langle\!\langle\ (\partial\, \mathfrak{a}\langle\, \text{x}_0\, \rangle\ |\ \mathfrak{b}\langle\, \text{x}_0\, \rangle) \otimes \partial_0\, \mathfrak{b}\langle\, \text{x}_0\, \rangle\ \rangle\!\rangle\ \rangle$$
$$(\partial\, \mathfrak{a}\langle\, \text{x}_0\, \rangle\ |\ \mathfrak{b}\langle\, \text{x}_0\, \rangle) \otimes (\partial_0\, \mathfrak{b}\langle\, \text{x}_0\, \rangle) \qquad \blacksquare$$

We wrote $[\ldots]$ above for the sake of brevity – the context of the congruence needs to be written out explicitly for Agda to be able to evaluate the metasubstitution that instantiates the distinguished hole metavariable. Note also the use of thm, which uses an established (non-axiomatic) equality as a proof step. The precise and sufficiently general definition of metasubstitution ensures that we always have access to the right metavariables and object variables where we need them, making the construction of equational proofs quite intuitive.

## 6 CONCLUSION

We presented a language-formalisation framework that allows users to produce Agda implementations of second-order languages at the press of a button. The generated term language is explicitly represented as an inductive, intrinsically-encoded data type, and the formalised metatheory can be used as and where required: substitution for operational semantics, compositional interpretations for denotational semantics, metasubstitution for equational reasoning, etc. All the formalisms featured in the library naturally derive from the mathematical theory of abstract syntax, without the need for ad-hoc definitions or lemmas. The convenient code-generation script allows for rapid prototyping and experimentation, and it is easy to manually extend a formalised signature or incorporate an existing intrinsic term syntax into the framework.

## 6.1 Related Work

The question of formalising and reasoning about abstract syntax was motivated by the development of proof assistants and the realisation that the Barendregt [1984] variable convention – *"rename variables as needed to avoid clashes"* – is difficult to translate into a formal setting. This has lead to a host of approaches that address the encoding of variable binding in proof assistants and functional languages, such as: higher-order abstract syntax [Pfenning and Elliot 1988; Chlipala 2008]; locally nameless representation [Bird and Paterson 1999; McBride and McKinna 2004; Weirich et al. 2011; Charguéraud 2012]; intrinsically-typed encoding [Benton et al. 2012; Allais et al. 2021; Érdi 2018]; and others [Shinwell et al. 2003; Urban and Kaliszyk 2011; Copello et al. 2017]. Similarly active is the mathematical study of abstract syntax and variable binding: developments include presheaf models [Fiore et al. 1999; Hofmann 1999]; nominal sets [Gabbay and Pitts 1999]; monadic approaches [Bellegarde and Hook 1994; Altenkirch and Reus 1999]; and others [Pigozzi and Salibra 1995; Sun 1999; Blanchette et al. 2019; Chen and Roşu 2020].

*Benchmarks.* The POPLMARK challenge [Aydemir et al. 2005; Abel et al. 2019] sets out a collection of criteria according to which metatheory-formalisation efforts can be compared. Several submissions use Coq as the target language, in some cases involving code generation from a second-order signature [Aydemir et al. 2008; Vouillon 2011; Lee et al. 2012; Polonowski 2013; Keuchel et al. 2016; Stark et al. 2019]. Though impressive, the approaches rarely adopt intrinsically-typed encodings of variables and terms, usually opting for numeric de Bruijn indices with all their complicated and error-prone arithmetic. We believe that the nameless, intrinsic representation is hard to surpass in dependently-typed proof assistants thanks to its static guarantees on the typing and scoping of terms. Its drawbacks (boilerplate and types "getting in the way") are significant and form one of the motivations of our line of research, but they are ultimately not unreasonable: a rigorous pen-and-paper proof of the type-preservation of substitution would involve the same difficulties we encounter in defining it in Agda. Our approach also incorporates generic traversals, and – for the first time, as far as we are aware – equational logic with the aid of parametrised metavariables, all naturally derived from the syntax.

*Type- and Scope-Safe Syntax.* Our work is closely aligned with that of AACMM [2021]: how to simplify the work of a language researcher by automating the boring metatheory. In their discussion of the presheaf model (loc. cit. Section 9.3), the authors suggest that freeing oneself from the formalities of the mathematics enables further progress in the development of the metatheory. We found very much the opposite: without the systematic view of the problem provided by the categorical model, one misses out on powerful principles that simplify proofs and untangle the conceptual labyrinth that is formal abstract syntax. A case in point is our reformulation of AACMM's core Semantics record as an instance of initial algebra semantics into the internal hom, which are very useful abstractions that the authors seemingly overlooked despite working from similar foundations. We also inherit the modularity of the categorical model by cleanly separating the mathematical groundwork, abstract metatheory, second-order signatures, and term representation; whereas AACMM's formalisation is very closely tied to their Desc data type which users are required to adopt to make use of the library. We do not have an analogue of their proof framework for simulations and fusions; but we suspect that equality-based properties (in addition to the fusion lemmas, which we already incorporate) can be captured through initiality, without reliance on another complex set of proof techniques.

*Presheaf Approach.* The backdrop of our Agda formalisation is a comprehensive mathematical reformulation of the presheaf model of variable binding [Fiore et al. 1999] with parametrised metavariables [Fiore 2008]. Namely, we shift attention to an indexed model (in the category of families over the set of contexts, equivalently functors from the discrete category $|\mathbb{F}|$ to **Set**) equipped with a pair of canonical adjoint monadic and comonadic modalities ($\Diamond \dashv \Box$, induced by the inclusion of $|\mathbb{F}|$ in $\mathbb{F}$) and their respective algebras and coalgebras, which are equivalent to presheaves. Although in this paper we have chosen not to emphasise the abstract categorical development in favour of a discussion geared towards programming-language researchers, we stress that the alternative viewpoint is not a matter of taste but crucial to the practical formal development that we have put forward. For instance, a direct translation of the presheaf model leads to a formalisation that inevitably requires quotients (colimits and coends) and cannot be used for computation (such as pretty printing, because of the lack of canonical representatives).

Our work required a variety of new considerations. Definitions and proofs had to be recast in the setting of (*i*) skew-closed structure and (*ii*) initiality; along the way, new notions and techniques had to be developed to bypass quotienting. Concerning (*i*), moving onto families with skew-monoidal structure is not enough to avoid the need for quotienting (e.g. Borthelle et al. [2020] achieve the freeness proof in a skew-monoidal setting only with the aid of a general lemma of Fiore and Saville [2017, Theorem 4.8] that relies on the presentation of initial algebras as colimits of $\omega$-chains, rather than as inductive data types); while, concerning (*ii*), the fact that initial algebras in presheaves can be lifted from initial algebras in families had to be given mathematical grounding. As an upshot, using families (indexed types), instead of presheaves, leads to a lightweight practical formalisation that suits the intrinsically-typed setting well.

## 6.2 Future Work

We recognise that the formal systems studied in modern type theory go far beyond second-order ones with algebraic types; indeed, linear, dual-context, polymorphic, dependent, polarised, etc. calculi abound. The presheaf approach has been extended to several of these [Tanaka 2000; Fiore 2006; Hyland and Tasson 2020; Fiore and Hamana 2013; Fiore 2008] and we are of course interested in adapting and/or extending our framework to these and combinations thereof.

As hinted at above, our background work involved a categorical reformulation of the highly abstract presheaf model to the formalisation-friendly realm of sorted families and $\Box$-coalgebras; indeed, much of the Agda code has been directly read off categorical definitions and commutative diagrams. Work on the formalisation, particularly in relation to metasubstitution, is ongoing. We also anticipate interesting applications in the study of parametrised signatures and signature translations; for example, we can encode the second-order equational theory of first-order logic and modularly extend it with the relation and function symbols of any first-order signature. Further experiments with more complex languages and proofs (such as the ones given in the PoplMark challenge) will also inform and motivate the future development of our library.

## ACKNOWLEDGMENTS

We thank the anonymous paper and artefact reviewers for their valuable suggestions and support. We are also grateful to Nathanael Arkor and Neel Krishnaswami for helpful discussions about technical and practical aspects of abstract syntax and language formalisation throughout the research.